\documentclass[12pt]{article}
\usepackage{amsmath}
\usepackage{graphicx}
\usepackage{enumerate, fourier}
\usepackage[inline, shortlabels]{enumitem}

\usepackage{natbib}
\usepackage{url} 
\usepackage[dvipsnames]{xcolor}
\usepackage{amsmath, amssymb, enumerate, graphicx, floatrow, subcaption, url, fancyhdr, dsfont, makecell}

\usepackage{times}
\usepackage{bm}

\usepackage{color}
\usepackage[left=1in,right=1in,top=1in,bottom=1in]{geometry}

\newtheorem{defn}{Definition}
\newcounter{assump}[section]
\newenvironment{assump}[1][]{\refstepcounter{assump}\par\medskip
	\noindent \textit{Assumption~\theassump. #1} \rmfamily}{\medskip}
\newtheorem{cond}{Condition}
\newcounter{thm}[section]
\newenvironment{thm}[1][]{\refstepcounter{thm}\par\medskip
	\noindent \textit{Theorem~\thethm. #1} \rmfamily}{\newline}

\newcounter{note}[section]
\newenvironment{note}[1][]{\refstepcounter{note}\par\medskip
	\noindent \textbf{Note~\thenote. #1} \rmfamily}{\medskip}

\newcounter{result}[section]
\newenvironment{result}[1][]{\refstepcounter{result}\par\medskip
	\noindent \textit{Result~\theresult. #1} \rmfamily}{\medskip}
\newcounter{lemma}[section]
\newenvironment{lemma}[1][]{\refstepcounter{lemma}\par\medskip
	\noindent \textit{Lemma~\thelemma. #1} \rmfamily}{\medskip}
\newcounter{remark}[section]
\newenvironment{remark}[1][]{\refstepcounter{remark}\par\medskip
	\noindent \textit{Remark~\theremark. #1} \rmfamily}{\medskip}
\numberwithin{equation}{section}
\begin{document}

\def\spacingset#1{\renewcommand{\baselinestretch}%
{#1}\small\normalsize} \spacingset{1}


  \title{\bf Controlling the False Discovery Rate in Complex Multi-Way Classified Hypotheses }
  \author{Shinjini Nandi\thanks{Department of Mathematical Sciences, Montana State University, USA. Email: shinjini.nandi@montana.edu}
    \hspace{.8cm}
         Sanat K. Sarkar\thanks{Department of Statistical Science, Temple University, USA. Email: sanat@temple.edu} \\
    }
  \maketitle
\bigskip
\begin{abstract}
In this article, we propose a generalized weighted version of the well-known Benjamini-Hochberg (BH) procedure. The rigorous weighting scheme used by our method enables it to encode structural information from simultaneous multi-way classification as well as hierarchical partitioning of hypotheses into groups, with provisions to accommodate overlapping groups. The method is proven to control the False Discovery Rate (FDR) when the p-values involved are Positively Regression Dependent on the Subset (PRDS) of null p-values. A data-adaptive version of the method is proposed. Simulations show that our proposed methods control FDR at desired level and are more powerful than existing comparable multiple testing procedures, when the p-values are independent or satisfy certain dependence conditions. We apply this data-adaptive method to analyze a neuro-imaging dataset and understand the impact of alcoholism on human brain. Neuro-imaging data typically have complex classification structure, which have not been fully utilized in subsequent inference by previously proposed multiple testing procedures. With a flexible weighting scheme, our method is poised to extract more information from the data and use it to perform a more informed and efficient test of the hypotheses. 
\end{abstract}

\noindent%

\spacingset{1.5} 
\section{Introduction}
\label{sec:intro}

In modern inferential problems arising from  large datasets, hypotheses often occur in complex structures of groups. Compelling examples can be found in neuro-imaging studies like functional magnetic resonance imaging (fMRI) or electroencephalography (EEG). Large sets of hypotheses obtained from such studies can be classified according to 
		\begin{enumerate*}[label = {(\alph*)}, series = tobecont, itemjoin = \quad]
		\item locations of interest (voxels in an fMRI study and electrodes in an EEG study) and\item their broader classifications (clusters of voxels known as Regions of Interest (ROIs) and regions of the cerebral cortex of the brain).
	\end{enumerate*}
The groups formed by the locations of interest can be nested within the groups formed by their broader classifications. In a different scenario, a set of hypotheses can be classified by more than one independent criteria that unlike the previous case, cannot be ordered among themselves. For example, \cite{STEIN20101160} conducted a study to detect voxelwise genomewide association in the human brain. The study involved the analysis of the relationship between $448293$ single-nucleotide polymorphisms(SNPs) and volume change in $31622$ voxels. 
The hypotheses obtained in this study can be classified according to groups formed by their SNPs and/or groups formed by a family for each voxel.
Furthermore, in such studies, when the grouping criteria are defined by the nature of the scientific experiment and statisticians have little or no control on their demarcation, it is quite likely the groups overlap with one or more hypotheses becoming members of more than one group. For example, in an EEG experiment, some electrodes are placed at the boundaries of two adjacent brain regions, thereupon, hypotheses that are associated to these electrodes can be classified as members of either brain regions. 

Though multiple hypotheses testing is a highly active research area, besides the p-filter algorithm proposed by \cite{ramdas2019}, not much development has been done to adequately justify overlapping groups, simultaneous or nested group structures in newly designed testing procedures. All the same, the grouping information provide valuable insight into the nature of the hypotheses which can be profitably used to create sharper distinction of signals from noise. 
The hypotheses classified into a group share common characteristics with other member hypotheses of the group. Consequently, a grouping structure levies a generic effect on its members, and a multiple testing method  that utilizes all such valuable information in its algorithm would be  naturally poised to make sharper distinction between the significant and non-significant state of an individual hypothesis. Some researchers have addressed the task of developing new multiple testing procedures adjusted to simple classification structure where hypotheses are classified into distinct groups by one norm of classification (\cite{Huetal2010}, \cite{BB}, \cite{Ignatiadis2016}, \cite{Sabha2019}, etc.). However, to the best of our knowledge, there does not exist a well-defined multiple testing procedure that controls the overall False Discovery Rate, by utilizing structural information gleaned from multiple classifications, that are either simultaneous, or hierarchical or both, and make provisions for overlapping groups. In the absence of such a procedure, an investigator has to prioritize a classification over the rest and choose a multiple testing procedure that accommodates non-overlapping groups formed by a single classification norm. In a situation where each of the multiple classifications contain equally important information, the investigator is obligated to compromise the accuracy of the analysis.

In this article we propose the theory of a generalized testing procedure, and its data-adaptive version, in an effort to alleviate this problem. We systematically develop a weighted Benjamini-Hochberg (BH)-type procedure, the Generalized Grouped BH (Gen-GBH) procedure from combining two weighted BH-type procedures, the Heir-GBH and the $S$-way GBH, designed for hierarchical classification and simultaneous classification structures respectively. Our proposed methods control the overall False Discovery Rate (FDR) in their respective classification settings. The theoretical formulations of the methods control FDR when the p-values satisfy some positive dependence criteria, while their data adaptive analogs control FDR under assumptions of independence. Due to the detailed classification information used in our approach, our proposed methods are poised to control FDR more accurately and possess more power than existing procedures that utilize partial or no classification information.

The paper is arranged as follows. In Section 2, we present a brief review of multiple testing research closely related to ours and some preliminary results at the base of our proposed methods. In Section 3, we present the p-value weighting schemes for hypotheses structured according to three different types of classification, hierarchical, multi-way simultaneous, and generalized classifications. The corresponding weighted BH methods in their oracle forms with proven FDR control under positive dependence and our proposals of their data-adaptive versions are also given in this section. The generalized classification integrates hierarchical classification into a simultaneous setting. 
Simulation studies in Section 4 investigate the performances of the oracle and data-adaptive weighted BH for hierarchically grouped hypotheses relative to their BH counterparts that ignore the underlying structure. Though the p-filter algorithm (\cite{ramdas2019}) considers a general classification setup such as ours, it serves FDR controlling objectives that are quite different from the objectives in this paper. Nevertheless, we include it in our comparative studies, noting the caveats caused by the differences in our goals. The results of these studies indicate that the p-value weighting scheme in our proposed weighted BH under hierarchical classification setting can capture the underlying structure quite effectively. This carries over to our proposed data-adaptive method under the generalized classification, as seen from its application to an EEG dataset, illustrated in Section 5. 
The paper concludes with Section 6 where some remarks are made on the effectiveness of multiple testing as an important statistical tool for analyzing complexly structured datasets. Proofs of some results associated with the FDR control of our proposed methods are given in the Appendix.

\section{Existing research and our contributions}
\subsubsection{Introduction of group structures in multiple hypotheses testing}
At the outset, multiple testing methods were designed to correct and secure multiplicity errors like the family-wise error rate and the false discovery rate, when testing a single set of hypotheses simultaneously. With their growing popularity as an inferential tool, their limitations in handling large datasets with low signal rates became evident (\cite{Pacifico2004}, \cite{meinshausen2009}). It was realized that clustering such hypotheses into local groups, and updating existing practices to incorporate the specific characteristics of the clusters formed leads to improved performance of these methods. The groups can be pre-specified, owing to the nature of the underlying scientific experiment, or can be formed based on domain specific knowledge, economic considerations, etc. The target is to utilize structural information about the groups formed and sharpen the contrast between the significant hypotheses and those that are not, leading to powerful and accurate decisions.

Grouping of hypotheses due to one criterion has become a common practice in multiple testing literature. We describe such classification due to one criterion as `one-way classification'. An early work by \cite{BenjHeller2007} discussed the importance of grouping hypotheses according to locations arising from a spatial data analysis. 
They argued that aggregation of hypotheses into local clusters tend to group similar hypotheses (nulls or non-nulls) into groups, thus increasing signal-to noise ratio.
\cite{Huetal2010} proposed an idea of using weighted p-values in the BH procedure where the weights capture the one-way classification structure. The weight assigned to each p-value depends on the composition of the group to which the p-value belongs to, and inflates or deflates the p-value conforming to the proportion of nulls in its parent group. The theoretical formulation of their Grouped BH (GBH) method controls the FDR at desired level of significance in finite samples. Two data-adaptive versions of the GBH method use p-value weights that are estimated from the data and control the FDR asymptotically when the p-values are independent or weakly dependent. 
In more recent work, 
\cite{Sabha2019} put forward the idea of structure adaptive BH algorithm (SABHA) that can incorporate structural information from one-way classified hypotheses in the form of weights. The weight $w$ assigned to each p-value is an estimate of the proportion of true nulls in its parent group, provided $w \in [\epsilon,1]$, for some arbitrary $\epsilon>0$. Such weights depend upon an estimate of the proportion of true nulls in the parent group of a p-value. The SABHA method controls FDR more liberally, and the excess over the level of significance is characterized by a term dependent on the Rademacher complexity of the estimated weights. They successfully apply the SABHA method to analyze an fMRI dataset, where hypotheses from measured voxels are grouped according to regions of interest. Substantial amount of research has been pursued to find powerful multiple testing procedures that are based on thresholding the Local False Discovery Rate (Lfdr, introduced by \cite{Efron2001}), applicable to one-way classified hypotheses. Few notable articles include \cite{caisun09}, \cite{Liu20161}, etc. The proposed methods in this paper do not utilize Lfdrs in their processes and are fundamentally different from the methods that depend on Lfdr thresholding.

Next we describe the development of our proposed Gen-GBH procedure in three stages, by combining the proposals of the Heir-GBH and the $S$-Way GBH, designed for hierarchical and simultaneous classification structures respectively. In its simple form, the oracle (theoretical) form of the Gen-GBH method conforms to the one-way classification structure as considered by all the testing procedures mentioned above. However, unlike any of the above methods, the Gen-GBH can accommodate overlapping groups in the one-way classification as well as in higher order classification structures. Hence it is capable of handling more complex group structures than one-way classified hypotheses. We show in section 3 that the oracle GBH procedure as proposed by \cite{Huetal2010} is a specific case of the oracle Gen-GBH. The data-adaptive version of the proposed procedure estimates the unknown parameters from the data and is theoretically guaranteed to control the FDR in finite samples, under the assumption that the p-values are independent. The SABHA method proposed by \cite{Sabha2019} is apt in handling one-way classified hypotheses in distinct groups. However, their choice of weights is quite different from our proposal and hence the control on FDR by the SABHA method, even in its oracle form is more liberal than our method, which, in its oracle form, attains exact control on FDR. 

\subsubsection{Our contribution}
Beyond one-way classification, there might be multiple classification norms that can be imposed concurrently on a set of hypotheses. A `hierarchical classification' is obtained when there is an ordering among the classification norms, implying that the groups formed by a certain norm can be split by a lower ordered norm to form smaller groups. A naturally occurring example is a phylogenetic tree, that classifies biological species into higher order groups, i.e., genera, which are further classified into higher order groups determined by the taxonomic hierarchy.
\begin{figure}
	\centering	
	\includegraphics[width=0.7\linewidth, height=0.5\textheight,keepaspectratio]{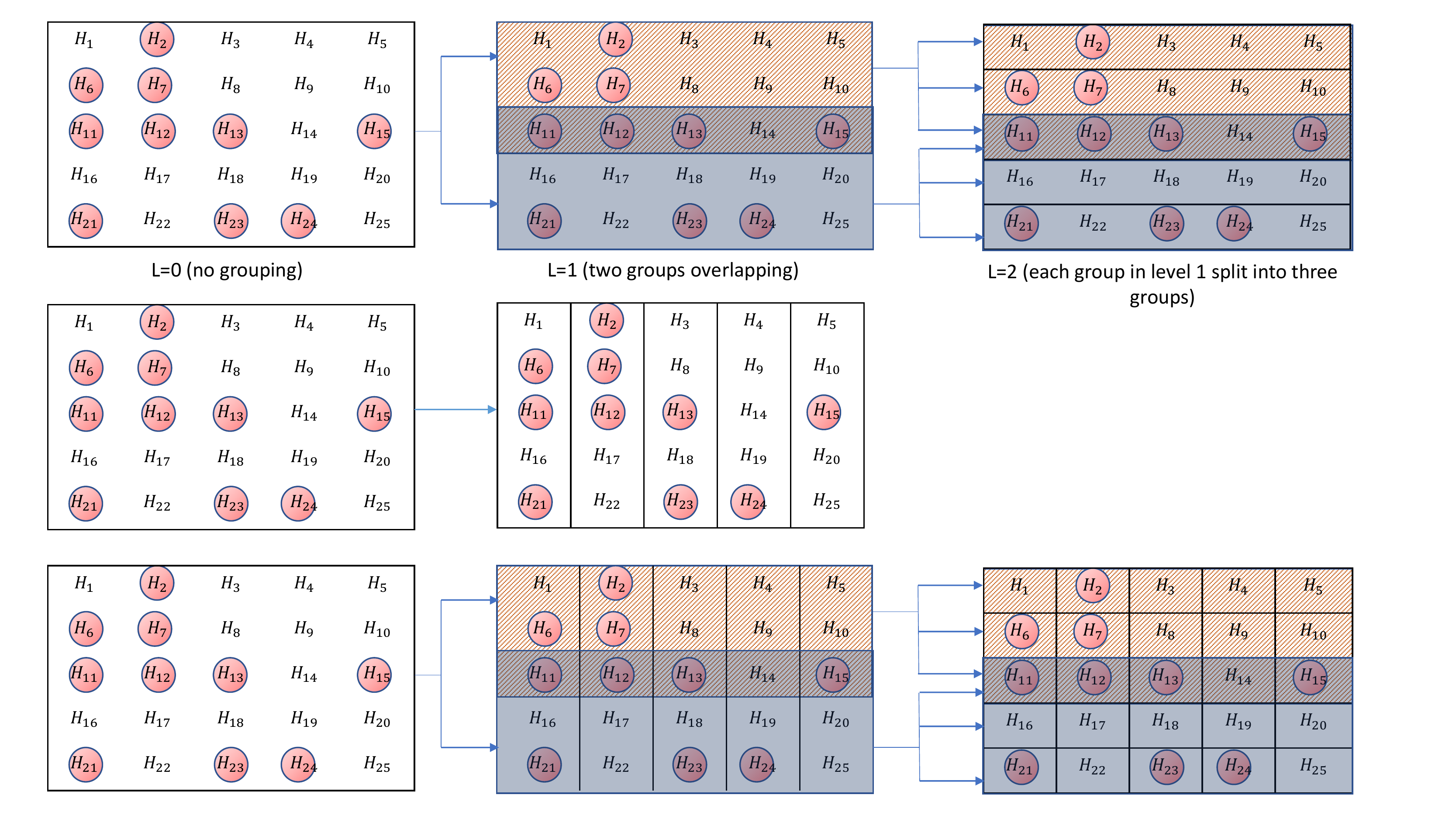}
	\caption{Figure showing the classification structures proposed in this article. 
	}
	\label{fighierarchical1}\end{figure}

To further illustrate the hierarchical structure, consider the following simple example. A set of
$N=25$ hypotheses is hierarchically classified in two levels. The top row in Figure \ref{fighierarchical1} shows the layout of the hypotheses and the classification structure in three stages.  The significant hypotheses are circled. Initially, at level L=0, there is no classification.  At level $L=1$, the set of $25$ hypotheses is split into two groups (shown in two different shades), each consisting of $15$ member hypotheses. There are five hypotheses at the intersection of the two groups. At level L=2, each of the two groups in level one are further split into three groups of five hypotheses each, each group being represented by a row. At any level, each hypothesis can be a member of an arbitrary number of groups, not exceeding the total number of groups at that level. The hierarchical procedures we propose in this paper are designed to account for the effect of all parent groups on each individual hypothesis. Theoretically, the hierarchical structure can be as detailed as desired, with groups at the ultimate level comprising of single hypotheses.

In a different situation, there may exist two or more classification criteria that may be imposed simultaneously, rather than hierarchically on a set of hypotheses. The simplest of such a case, involving two simultaneous criteria giving rise to `two-way classified' hypotheses was discussed in a recent article by \cite{NANDI2021}. 
A common example of simultaneously classified hypotheses can arise from any spatio-temporal data, where the spatial classification norm like geographical regions, brain voxels, etc. is unrelated to the temporal norm like seconds, hours, days, etc., but can be simultaneously imposed on the same set of hypotheses.
We seek to combine the idea of hierarchical and simultaneous classifications to suggest a generalized classification setup, that also allows for overlapping groups formed by one or multiple criteria. In Figure \ref{fighierarchical1}, the middle row shows that the same set of $N$ hypotheses can be classified according to a different norm along the columns. The hierarchical classification along rows and the classification along the columns can be combined, as shown in the bottom row of Figure \ref{fighierarchical1}, to form the generalized classification setup.
Rest of this paper is devoted to the development and discussion of the generalized weighted multiple testing procedures 
to account for hierarchical and overlapping groups that may occur along the simultaneous classifications, and control the False Discovery Rate. 

\subsubsection{Critical appraisal of our method in comparison to other recent methods} 
P-value weighting is a well-established concept to boost power of the multiple testing procedures. The weights can incorporate structural information (earliest example of which can be found in \cite{Storeyetal2004}, \cite{Benjaminietal2006}, \cite{Blanchard:2009:AFD:1577069.1755880}, \cite{Sarkar2008},etc.), penalties for multiple decisions (\cite{Benjamini1997}, etc.). 
Some articles like \cite{Genov2006} and \cite{Ignatiadis2016} propose formulation of weights that do not depend directly on the p-values, but use prior information about the experiment and external covariates to reflect upon the nature of the p-values. Various formulations of complex structures of p-values, have been discussed in the literature, along with the novel designs of multiple testing procedures to address such structures. For example, in a `tree structure' of hypotheses, a p-value is tested only if its parent is rejected. Several methods including \cite{yekutieli2006}, \cite{yekutieli2008}, \cite{star2020}, etc. address such structures of hypotheses. A different setup of `ordered hypotheses' and corresponding testing procedures have been discussed in \cite{LeiFithian2016}, \cite{Sabha2019}, etc. The ordered structure of hypotheses assumes that hypotheses early in a lineup are more likely to contain true signals than their successors. Though our proposed methods address hierarchically classified hypotheses, clearly such classification is not to be confused with tree structures or ordered hypotheses. 
Hence much as our proposed methods are not suited for the structures mentioned above, to the best of our knowledge, the methods that address tree structures and ordered structures are not suited to the setup we discuss here. 

To the best of our knowledge, the p-filter algorithm developed in \cite{2015arXiv151203397F} and \cite{ramdas2019} is the only algorithm, besides ours, able to address hierarchical and/or simultaneous classifications with presence of overlapping groups. However, a basic difference between the p-filter and our algorithms is that the former controls FDR in each group and level of every classification, consequently leading to quite a conservative control on the overall FDR. While controlling FDR at each type of classification is desirable in some scenarios, our goal is to control the overall FDR and enhance the accuracy of the BH procedure using the classification information. Consequently, our methods are less conservative, and more accurately control the FDR than the p-filter process. 

\subsection{Preliminaries}

\subsubsection{The Weighted BH procedure.}
We begin the discussion from some fundamental results on multiple testing in the framework of controlling the False Discovery Rate (FDR), as defined by \cite{BH95}. Some results from \cite{NANDI2021} are re-stated and generalized to form the background of the results obtained in this article.
The goal of a multiple testing procedure is to discover the false null hypotheses with a control on the number of false discoveries. Consider a set of hypotheses $H_1, \ldots, H_N$ that are to be tested simultaneously. 
The False Discovery Rate (FDR) is the expected proportion of false discoveries and is defined as 
\begin{align*}
	\text{FDR} = E\left[\frac{V_N}{\max(R_N, 1)}\right],
\end{align*}
with $V_N$ and $R_N$ being the number of false and total discoveries made, respectively.
The following assumptions are made on the nature of the p-values.
\begin{assump}\label{assump1}
	Let $I^0 \subseteq \{1, \ldots, N \}$ be the set of indices of the true null hypotheses. For each $i \in I^0$, $P_i \sim \text{Uniform}(0,1)$.
\end{assump}
\begin{assump}\label{assump2}
	The set of p-values $P_1, \ldots, P_N$, corresponding to $H_1, \ldots, H_N$, are Positively Regression Dependent on the Subset (PRDS) of null p-values.
\end{assump}

A set of random variables $X_1, \ldots, X_k$ is said to be positively regression dependent on a subset $S$ of these random variables if $E\left[\phi(X_1, \ldots, X_k)| X_i = x\right]$ is non-decreasing in $x$, for each
$X_i \in S$ and for any (coordinatewise) non-decreasing function $\phi$ of $X_1, \ldots, X_k$. 
A weaker form of positive dependence property, with $E\left[\phi(X_1, \ldots, X_k)| X_i = x\right]$ replaced by $E\left[\phi(X_1, \ldots, X_k)| X_i \le x\right]$, is often assumed in the literature in the context of
BH-type FDR controlling procedures \citep{finner2009,Sarkar2008}. Our proposed methods provably control the FDR under either of these conditions defining the PRDS property.

A general form of the Benjamini-Hochberg (BH) procedure 
forms the building block for the multiple testing procedures we develop in this article.

\begin{defn}\label{defn1}{\rm [Weighted BH]} \;
	Given non-stochastic weight $W_i \ge 0$ assigned to $P_i$, for $i=1, \ldots, N$, the weighted BH at level $\alpha$ is a level $\alpha$ BH method applied to the weighted p-values $P_i^W = W_iP_i, i=1, \ldots, N$, i.e., it orders the weighted p-values as $P_{(1)}^W \le \cdots \le
	P_{(N)}^W$, and rejects the hypothesis $H_{(i)}$ corresponding to $P_{(i)}^W$, for all $i=1, \ldots, \max \left \{1\leq j \leq N: P_{(j)}^W \leq j\alpha/N \right \}$, provided the maximum exists; otherwise, it rejects
	none.
\end{defn}
\begin{result}\label{result1}
	For p-values satisfying the PRDS property, the FDR of the weighted BH is bounded above by $\alpha \sum_{i \in I_0}W_i^{-1}/N$.
\end{result}

The following condition ensures that the weighted BH controls FDR at level $\alpha$.
\begin{cond}\label{cond1}
	The weights $W_i$, assigned to the p-values $P_i, i=1, \ldots, N$, are such that $\sum\limits_{i \in I^0} W^{-1}_i = N$.
\end{cond}
Result \ref{result1} is re-stated from \cite{NANDI2021} and interested readers are referred to it for the proof. 
Condition 1 permits researchers to conveniently incorporate a wide range of structural information about the hypotheses in the
weights. The weighted BH thus encompasses several multiple testing procedures that employ any formulation of weights that satisfy Condition 1. 
Notably, the classical BH 
corresponds to the case when $W_i= \pi^0$, the proportion of true nulls, for all
$i=1, \ldots, N$. 

\subsubsection{Moulding the Weighted BH procedure for group structures with overlapping groups}
Prior to applying the Weighted BH method to complex group structures, it is important to note that the weights $W_i$ can be designed to allow overlapping groups formed by a classification norm. Consider a situation where a set of $N$ hypotheses are classified into $m (> 1)$ groups, and each group can overlap with one or more groups. The following result shows how the weights can be formulated to capture this structure.

\begin{result}\label{result2} Given any $w_g > 0$,
	$g = 1, \ldots, m$, quantifying the underlying structural information from the $g$th group, let the $i$th p-value be assigned the weight $W_i$ given below:
	\begin {eqnarray} \frac{1}{W_i} = \left (\frac {1}{N}
	\sum_{g=1}^m \frac{n_g \pi_g^0}{w_g} \right )^{-1} \sum_{g':I_{g'} \ni i} \frac{1}{w_{g'}},\qquad i=1, \ldots, N,\end{eqnarray}  where $n_g$ is number of hypotheses and $\pi^0_g$ is the proportion of true hypotheses in group $g$. These weights satisfy Condition \ref{cond1}.

\end{result}

Result \ref{result2} is proved in the Appendix. This scheme of weighting generalizes the one-way classification structure of hypotheses with non-intersecting groups 
and lays the foundation to design suitable multiple testing procedures for overlapping groups in more complicated structures of hypotheses.


\subsubsection{Data-Adaptive Version of the Weighted BH procedure.}
\cite{Storey2002} and \cite{Storeyetal2004} proposed a point estimate of the number of true nulls in a given set of independent p-values. This estimate, based on a tuning parameter $\lambda$ was advantageously
used in \cite{Storey2002} and \cite{Storeyetal2004} and in subsequent literature (\cite{Blanchard:2009:AFD:1577069.1755880}, \cite{Sarkar2008}, etc.) to propose new estimates of the proportion of true nulls. These
estimates, when substituted for weights in the Weighted BH procedure, helped design data-adaptive versions of the method that controls FDR when p-values are independent. We extend this idea to estimate the weights in our proposed hierarchical and/or simultaneous classification of hypotheses, such that they satisfy the following result.
\begin{result}\label{result3} For a data-adaptive weighted BH procedure with (coordinatewise) non-decreasing estimated weight functions $\hat{w}_i(\mathbf{P}) > 0$, $i=1,\ldots, N$, FDR $\le \alpha$ under independence if \begin{align}\label{e2.2}
	E\left \{\sum\nolimits_{i \in I_0} \left [1/\hat{w}_i(\mathbf{P}^{(-i)},0) \right] \right \} \leq N, \end{align} where $\hat{w}_i(\mathbf{P}^{(-i)},0)$ represents $\hat{w}_i$ as a function of $\mathbf{P}^{(-i)} = \{P_1, \ldots, P_N\}\setminus \{P_i\}$ with $P_i = 0$.
\end{result}

Interested readers are referred to \cite{Sarkar2008} for a proof of this result.
\section{Proposed multiple testing methodologies}
\subsection{Hierarchical Classification}\label{sec_hier_class}  Suppose there are $N$ hypotheses that can be partitioned into groups according to each of $L (\ge 1)$ different classification criteria. Assuming that these criteria are intrinsically ordered from $1$ to $L$, we consider imposing all these classifications hierarchically. At level $l$, i.e., upon successive application of the first $l$ classifications, each of the groups can be further partitioned into at least one subgroup at level $l+1$, where $l=0,1, \ldots L-1$. When $l=0$, we have the unclassified group of all $N$ hypotheses, and as $l$ increases, we have non-decreasing
number of groups with non-increasing number of member hypotheses in each group.
At any level $(\ge 1)$, there might be two or more groups that can overlap, i.e., they have one or more hypotheses in common.

Let $G(g_1\cdots g_l)$, $g_l= 1, \ldots, m_l(G(g_1\cdots g_{l-1}))$, be the $m_l(G(g_1\cdots g_{l-1}))$ subgroups of hypotheses formed at level $l$ from the $(g_1\ldots g_{l-1})$th subgroup $G(g_1\cdots g_{l-1})$, for $l=0,1, \ldots, L$, with $G(g_0)$ referring to
the original unclassified set of $N=m_0$ null hypotheses. For notational convenience, henceforth, we refer to $m_l(G(g_1\cdots g_{l-1}))$ as $m_l$.
Thus, the hierarchical classification scheme partitions the $N$ hypotheses into the $\prod_{l=1}^L m_l$ subgroups $G(g_1\cdots g_L)$, $(g_1, \ldots g_L) \in \prod_{l=1}^L \{1, \ldots, m_l\}$.

Let $I_{g_1\cdots g_l}$ and $I^{0}_{g_1\cdots g_l}$, respectively, be the sets of indices of the nulls and true nulls in $G(g_1\cdots g_l)$, with  $n_{g_1\cdots g_l} = |I_{g_1\cdots g_l}|$, $n^{0}_{g_1\cdots g_l} =
|I^{0}_{g_1\cdots g_l}|$, and ${\pi}^{0}_{g_1\cdots g_l} = n^{0}_{g_1\cdots g_l}/ n_{g_1\cdots g_l}$, being the number of nulls (which is $N$ when $l=0$), number of true nulls, and the proportion of true nulls,
respectively, in this group. The proportion of true nulls in the entire set of null hypotheses, at any level $l$, is $\pi^0 = |I^0|/N$, where $I^0 = |\bigcup_{g_1=1}^{m_1} \cdots \bigcup_{g_l=1}^{m_l} I^{0}_{g_1\cdots
g_l}|$ and $N = |\bigcup_{g_1=1}^{m_1} \cdots \bigcup_{g_l=1}^{m_l} I_{g_1\cdots g_l}|$, which reduce, respectively, to \\ $|I^0| = \sum_{g_1=1}^{m_1} \cdots \sum_{g_l=1}^{m_l} n_{g_1\cdots g_l} \pi^{0}_{g_1 \cdots g_l}$ and
$N=\sum_{g_1=1}^{m_1} \cdots \sum_{g_l=1}^{m_l} n_{g_1\cdots g_l}$ when no overlapping occurs at any level of classification.




\subsubsection{Oracle Hierarchically Grouped BH (Heir-GBH)}
Assuming ${\pi}^{0}_{g_1\cdots g_l}$ to be known for all $(g_1, \ldots, g_l)$ and $l$, we will now construct the desired weights $W_{i}$ satisfying Condition 1, i.e.,
\begin{eqnarray} \label{e3.1}\sum_{i \in I^0} \frac{1}{W_i}\mathds{1}\left (i \in \bigcup_{g_1=1}^{m_1} \cdots \bigcup_{g_L=1}^{m_L} I_{g_1 \cdots g_L} \right ) =  N. \end{eqnarray}
Our proposed procedure assigns the weight $W_i$ to each $P_i \in \bigcup_{g_1=1}^{m_1} \cdots \bigcup_{g_L=1}^{m_L} G(g_1 \cdots g_L)$, and applies the BH method to these weighted $p$-values.

\begin{lemma}\label{lemma1}
Let $W_i$ be such that
\begin{eqnarray} \label{e3.2} \qquad \frac{1}{W_i} = \left ( \frac{1}{N} \sum_{g_1=1}^{m_1} \cdots \sum_{g_L=1}^{m_L} \frac{n_{g_1 \cdots g_L}^0}{w_{g_1 \cdots g_L} }\right )^{-1} \sum_{g_1=1}^{m_1} \cdots
	\sum_{g_L=1}^{m_L} \frac{\mathds{1} \left ( I_{g_1 \cdots g_L} \ni i \right )}{w_{g_1 \cdots g_L}}, \end{eqnarray} where $w_{g_1 \cdots g_L}$ is recursively defined below:
\begin{align}\label{e3.3}
	w_{g_1 \cdots g_L}  = \frac{\pi^0 (1- \pi^0)}{w_{g_1 \cdots g_{L-1}}} \frac{\pi_{g_1 \cdots g_{L}}^0}{1-\pi_{g_1 \cdots g_{L}}^0},  \end{align}
with $w_{g_0}= \pi^0$. The $W_i$'s in (\ref{e3.2}) satisfy Condition 1.  \end{lemma}

\begin {note}
There is a variety of choice for $w_{g_1 \cdots g_L}$ to define $W_i$ satisfying Condition 1, as suggested by Result \ref{result2}. For instance, one can simply choose $w_{g_1 \cdots g_L} = \pi_{g_1 \cdots
g_{L}}^0/(1-\pi_{g_1 \cdots g_{L}}^0)$. Such a choice, however, does not capture the underlying hierarchical nature of the successive
groupings, unlike our chosen $w_{g_1 \cdots g_L}$ in Lemma \ref{lemma1}. The $W_i$ in Lemma \ref{lemma1} defined through this $w_{g_1 \cdots g_L}$ accounts for the effects of all the parent partitions that contain the hypothesis $H_i$
corresponding to $P_i$. It sequentially compares the significance of composition of each group containing $H_i$ at a particular level with all other groups originating from the same parent, thus recording the impact of
the entire hierarchical setup on the individual hypotheses.

It is important to note that if overlapping does not occur at any of the $L$ levels, then $W_{i} = w_{g_1 \cdots g_L}$, for each $P_i \in G(g_1 \cdots g_L)$, where $(g_1, \ldots, g_L) \in \prod_{l=1}^L \{1, \ldots,
m_l\}$.

The following alternative expression for $w_{g_1 \cdots g_L}$ is worth noting.
\begin{align}\label{e3.4}
w_{g_1 \cdots g_L}  = w_{g_1 \cdots g_{L-2}} \frac{1-\pi_{g_1 \cdots g_{L-1}}^0}{\pi_{g_1 \cdots g_{L-1}}^0}\frac{\pi_{g_1 \cdots g_{L}}^0}{1-\pi_{g_1 \cdots g_{L}}^0},  \end{align}
with $w_{g_0}= \pi^0$ and $w_{g_1}= (1-\pi^0)\pi_{g_1}^0/(1- \pi_{g_1}^0)$. This will be used to show that $w_{g_1 \cdots g_L}$ satisfies Condition 1 while proving Lemma \ref{lemma1} in the Appendix.
\end{note}

\begin{defn}[Oracle Hierarchically Grouped BH (Heir-GBH)] The level $\alpha$ BH procedure applied to $W_iP_i$ for all $P_i \in \bigcup_{g_1=1}^{m_1} \cdots \bigcup_{g_L=1}^{m_L} G(g_1 \cdots g_L)$, where $W_i$ is
defined in (\ref{e3.2}) and (\ref{e3.3}). \end{defn}
\begin{thm} The Heir-GBH controls the FDR under Assumptions \ref{assump1} and \ref{assump2}. \end{thm}
This theorem follows from Result 1 and Lemma 1.
\begin{note} When $L=0$, i.e., when no classification is applied, the Heir-GBH assigns the weight $w_{g_0} = \pi^0$ to each p-value, which is the same weight assigned by the Oracle BH. to all p-values. 
\end{note}
\begin{note}
When $L=1$,
$w_{g_1} =  (1 - \pi^0) \cdot \pi_{g_1}^0\big/(1-\pi_{g_1}^0)$, for $g_1=1, \ldots m_1$. If there are no overlapping groups,
the $i$th p-value of group $G(g_1)$ is assigned the weight $W_i = w_{g_1}$. The Heir-GBH for $L=1$, thus reduces to the Oracle One-Way GBH in \cite{Huetal2010} and ~\cite{NANDI2021} which address one-way classified hypotheses with non-overlapping groups. 

\end{note}

\subsubsection{Data-Adaptive Procedure} The data-adaptive version of Heir-GBH relies on the data to estimate the weights assigned to the p-values. We generalize the novel idea of estimating such weights for one-way
classified hypotheses proposed in \cite{NANDI2021}, and formulate the estimated weights in the
current setting of more complex hierarchically grouped structure of hypotheses.

Given a specific $\lambda \in (0,1)$, the number of p-values below $\lambda$ is determined at the last level $L$ as $R_{g_1 \cdots g_L}(\lambda)  = \sum_{i \in I_{g_1 \cdots g_L}} \mathds{1}(P_i \le\lambda),$ and the estimate of ${n}^0_{g_1 \cdots g_L}$ as 
\begin{align}\label{eqn7}
	\hat{n}^0_{g_1 \cdots g_L} =  \frac{n_{g_1 \cdots g_L}-R_{g_1 \cdots g_L}(\lambda) +1}{1-\lambda}
\end{align}
On the basis of $\hat{n}^0_{g_1 \cdots g_L}$, for a generic group $G(g_1 \cdots g_l)$ at level $l$, we successively define the estimate of $n^0_{g_1 \cdots g_l}$ as
\begin{align}\label{eqn8}
\hat{n}^0_{g_1 \cdots g_{l-1}} =  m_l\cdot \hat{n}^0_{g_1 \cdots g_l} \;\forall \;l =1, \ldots, L.
\end{align}
and the proportion of true nulls ${\pi}^0_{g_1 \cdots g_l}$, as  
$
	\hat{\pi}^0_{g_1 \cdots g_l} = \frac{\hat{n}^0_{g_1 \cdots g_l}}{n_{g_1 \cdots g_l}}, \forall \;l =0, \ldots, L.
$

If $L=0$, i.e., for a single group of hypotheses, the estimated weight $\widehat{W}_i$ assigned to the $i$th p-value is  $(N-R_N(\lambda)+1)/N(1-\lambda)$, where $R_N(\lambda) = \sum_{i=1}^N \mathds{1}(P_i \le \lambda)$ and that at level 
$L(\ge 1)$ is 
\begin{eqnarray} \label{e3.6} \qquad \frac{1}{\widehat{W}_i} = \left ( \frac{1}{N} \sum_{g_1=1}^{m_1} \cdots \sum_{g_L=1}^{m_L} \frac{\hat{n}_{g_1 \cdots g_L}^0}{\hat{w}_{g_1 \cdots g_L} }\right )^{-1} \sum_{g_1=1}^{m_1}
\cdots \sum_{g_L=1}^{m_L} \frac{\mathds{1} \left ( I_{g_1 \cdots g_L} \ni i \right )}{\hat{w}_{g_1 \cdots g_L}}, \end{eqnarray} where
$\widehat{w}_{g_1\cdots g_L}$ estimates the grouping effect of $G(g_1 \cdots g_L)$, and is given by the recurring formula
\begin{eqnarray}\label{e3.7}&
\hat{w}_{g_1 \cdots g_L} = \left \{
\begin{array}{ll}
\frac{\hat{n}_{g_1}^0}{N}\cdot m_1 & \mbox{ for}\; L =1 \\
\hat{w}_{g_1 \cdots g_{L-2}}\frac{\hat{n}_{g_1 \cdots g_L}^0}{\hat{n}_{g_1 \cdots g_{L-1}}^0}\cdot m_L & \; \mbox{for} \; L \ge 2, \\ 
\end{array}
\right.
\end{eqnarray}
where $\hat{w}_{0} = \hat{\pi}^0$. 
This leads us to propose the data-adaptive multiple testing procedure for hierarchically classified hypotheses as given below:

\begin{defn} [Data Adaptive Hierarchically Grouped BH (DAHeir-GBH)] The level $\alpha$ BH procedure applied to $\widehat{W}_iP_i$ for all $P_i \in \bigcup_{g_1=1}^{m_1} \cdots \bigcup_{g_L=1}^{m_L} G(g_1 \cdots g_L)$,
where $\widehat{W}_i$ is given by (\ref{e3.6}) and (\ref{e3.7}).\end{defn}

\begin{note}
Like Heir-GBH, DAHeir-GBH reduces to a method that uses $\widehat{W}_i = \widehat {w}_{g_1 \cdots g_L}$ for each $P_i \in G(g_1 \cdots g_L)$, if no groups overlap at any of the $L$ levels.
\end{note}
\begin{thm}\label{theorem2}
The DAHeir-GBH controls FDR at level $\alpha$, under the assumption that the p-values are independent.
\end{thm}
The proof of this theorem is provided in the Appendix.

\subsection{Simultaneous Multi-Way Classification}\label{sec_sim_class}
Suppose there exist $S$ ($\ge 2)$ different classification criteria, each of which can simultaneously be used to split the set of $N$ hypotheses into groups, thus resulting in an $S$-way classification structure for the
hypotheses. In the simple case of $S=2$, i.e., for two-way classification (considered in \cite{NANDI2021}, with non-overlapping groups for each classification), the hypotheses can be arranged in the form of an $M_1 \times M_2$
matrix. The $M_1$ rows and $M_2$ columns of the matrix represent two sets of non-overlapping groups created out of the $N$ hypotheses by the two classifications. For a general $S$,
this structure resembles an $S$-dimensional array, with the $s$th array representing a set of $M_s$ groups, created out of the $N$ hypotheses by classification $s$.

Although the groups formed by each classification can be allowed to overlap, we defer discussion of this case to the next sub-section where a generalized classification structure is considered by integrating the ideas of
hierarchical and simultaneous groupings.

Let $g_s$ be the $g$th group containing $n_{g_s}$ hypotheses for classification $s$, for $g_s= 1, \ldots, M_s,\; s=1, \ldots, S$.
Then, with $I^0_{g_s}$ denoting the subset of indices of true nulls in group $g_s$, $\pi^0_{g_s} = \lvert I^0_{g_s} \rvert / n_{g_s}$ is the proportion of true nulls in group $g_s$ for classification $s$, $s=1, \ldots,
S$, and
\begin{align}
\pi^0 = \frac{\sum\limits_{g_s=1}^{M_s} \lvert I^0_{g_s} \rvert}{\sum\limits_{g_s=1}^{M_s}n_{g_s}} = \frac{\sum\limits_{g_s=1}^{M_s} n_{g_s} \pi^0_{g_s}}{\sum\limits_{g_s=1}^{M_s}n_{g_s}}
\end{align} is the proportion of true nulls in the entire set of $N$ hypotheses for any of these classifications.

The simultaneous classification schemes are assumed to act independently of each other, so the groups of hypotheses do not overlap between classifications, except at the elementary level where each hypothesis is
classified concurrently in $S$ different ways. In other words, each $P_i$ can be re-indexed as $P_{g_1 \cdots g_S}$, i.e., can be identified as the p-value appearing at the intersection of the $S$ groups $g_1, \ldots,
g_S$, where $(g_1, \ldots, g_S) \in \prod_{s=1}^S \{1, \ldots, M_s\}$. With this in mind, we proceed to define the weight to be attached to each $P_i$ and propose our multiple testing procedure in its oracle form under $S$-way classified setting in the following sub-section.

\subsubsection{Oracle Procedure}
Assuming $\pi^0_{g_s}$ to be known, for all $g_s = 1, \ldots, M_s$, $s=1, \ldots, S$, we will construct the desired weights $W_i$
satisfying Condition 1, i.e.,
\begin {eqnarray} \label{e3.9}  \sum_{i \in I^0} \frac{1}{W_i} \sum_{g_1=1}^{M_1} \cdots \sum_{g_S = 1}^{M_S} \mathds{1} \left ( (g_1 \cdots g_S) = i \right ) = N, \end {eqnarray} The weight $W_i$ will be attached to
$P_{i}$, and the BH method applied to these weighted $p$-values will be our proposed method in its oracle form under $S$-way classified setting of the hypotheses.

The following lemma gives the desired weights.

\begin{lemma}\label{lemma2}
Let $W_i = w_{g_1 \cdots g_S}$ when $i=(g_1 \cdots g_S)$, where
\begin{align}\label{e3.10}
\frac{1}{w_{g_1 \cdots g_S}}  = \frac{1}{S} \sum_{s=1}^S \frac{1}{w_{g_s}}, \; \mbox{and} \; w_{g_s} = \frac{\pi^0_{g_s}(1-\pi^0)}{1-\pi^0_{g_s}},  \end{align}
for $(g_1, \ldots, g_S) \in \prod_{s=1}^S \{1, \ldots, M_s\}$. These $W_i$'s satisfy Condition \ref{cond1}.  \end{lemma}

This lemma follows from the fact that the left-hand side of (\ref{e3.9}) with the $W_i$ in the lemma equals $\frac{1}{S} \sum_{s=1}^S\sum_{i \in I^0_{g_s}}\frac{1}{w_{g_s}}$, which is $N$, since $\sum_{i \in I^0_{g_s}}\frac{1}{w_{g_s}} = N$, for each $s$.

\begin {note} This weight assigned to $P_i$ is the simple mean of the effects imposed on it by all its $S$ parent groups. It accounts for the significant effect of each group that this p-value belongs to. In case any of
the $S$ classification norms is ineffective or does not provide information substantial to distinguish between the states (significant or not) of the member p-values, the term related to its effect can be easily removed
from the formula (\ref{e3.10}), and the weight can be recalculated to consider the effects of the remaining $S-1$ grouping norms. \end {note}

We are now ready to define our proposed multiple testing procedure in its oracle form for simultaneously $S$-way classified hypotheses.

\begin{defn}{\bf Oracle $S$-Way Grouped BH ($S$-Way GBH)} is defined as the level $\alpha$ BH procedure applied to $W_iP_i$, for $i=1, \ldots, N$, where  $W_{i}$ is defined as in (\ref{e3.10}). \end{defn}

\begin{thm}\label{theorem3} The $S$-Way GBH controls the FDR under Assumptions \ref{assump1} and \ref{assump2}. \end{thm}
This theorem follows from Result 1 and Lemma \ref{lemma2}.

\subsubsection{Data-Adaptive Procedure} Similar to the way data-adaptive weights are chosen to define DAHeir-GBH, 
we consider estimating the weight $W_i$ for $P_i$ in developing the
data-adaptive form of Oracle $S$-Way GBH as follows:
Let \begin{align}\label{e3.11}
\widehat{W}_i = \hat{w}_{g_1 \cdots g_S} \; \mbox{when} \; i=(g_1 \cdots g_S), \end{align} 
\begin{align} \label{e3.12} \text{where }\qquad \frac{1}{\hat{w}_{g_1 \cdots g_S}}  = \frac{1}{S} \sum_{s=1}^S \frac{1}{\hat{w}_{g_s}}, \; \mbox{and} \; \hat{w}_{g_s} = \frac{n_{g_s} - R_{n_{g_s}}(\lambda) + 1}{N(1-\lambda)}\cdot M_s, 
\end{align} with $R_{n_{g_s}}(\lambda)$ representing 
the number of p-values not exceeding $\lambda$ in group $g_s$. 
Based on these weights, we propose our data-adaptive procedure under simultaneous classification setting as follows:
\begin{defn} {\bf Data Adaptive $S$-Way Grouped BH (DA-$S$-Way GBH)}  is defined to be the level $\alpha$ BH procedure applied to $\widehat{W}_iP_i$, for $i=1, \ldots, N$, where  $\widehat{W}_{i}$ is defined as in (\ref{e3.11}) and (\ref{e3.12}). \end{defn}

\begin{thm}\label{theorem4}
The DA-$S$-Way GBH controls FDR at level $\alpha$, under the assumption that the p-values are independent.
\end{thm}
This theorem is proved in the Appendix.
\subsection{Generalized Classification}\label{section3.3}

The generalized setup allows the grouping structures to be more detailed. It accommodates hierarchical and simultaneous classifications concurrently, combining the ideas from Sections \ref{sec_hier_class} and \ref{sec_sim_class}. More specifically,
the set of $N$ hypotheses can be subjected to simultaneous (one-way) classifications according to $S$ different criteria, with the $s$th criterion allowing hierarchical groupings, possibly overlapping, using $L_s$ ($\ge
2$) ordered grouping norms, as described in Section \ref{sec_hier_class}.

Integrating the notations used for hierarchical classifications into the simultaneous setting, let us define $G(s; g_1 \cdots g_{l})$ as the $(g_1 \cdots g_{l})$th group created out of the $N$ hypotheses by applying the
hierarchical classification scheme up to the first $l$ levels at the $s$th simultaneous classification, where $(g_1, \ldots g_l) \in \prod_{l=1}^{L_s}\{1, \ldots, m_{l}\}$, $s=1, \ldots, S$.

\subsubsection{Oracle Procedure} Assuming the proportion of true nulls, $\pi^0_{s;g_1 \cdots g_l}$, in $G(s; g_1 \cdots g_{l})$, to be known for all $l=1, \ldots, L_s$, $s=1, \ldots, S$, we want to construct the weight
$W_i$ to be attached to $P_i$. This would pave the way toward defining our proposed procedure under the generalized $S$-way classification. The following lemma gives the desired $W_i$.
\begin{lemma}\label{lemma3}  Let $W_i$ be such that
\begin {eqnarray} \label{e3.13}\frac{1}{W_i} = \frac{1}{S} \sum_{s=1}^S \frac {1}{W_i(s)}, \end{eqnarray} where, for each $s$, $W_i(s)$ is defined as in Lemma 2 in terms of $\pi^0_{s;g_1 \cdots g_l}$, $l=1, \ldots, L_s$. Such $W_i$'s, for all $i = 1, \ldots, N$ satisfy Condition \ref{cond1}.
\end{lemma}

The lemma can be easily proved by noting from Lemma \ref{lemma1} that $\sum_{i \in I^0} \frac{1}{W_i(s)} = N$, for each $s$.

Our proposed procedure in its oracle form, under generalized multi-way classification setting, is formally defined in the following:
\begin{defn} The {\bf Oracle Generalized Grouped BH (Gen-GBH)} is a level $\alpha$ BH procedure applied to $W_iP_i$, for $i=1, \ldots, N$, where  $W_{i}$ is defined as in (\ref{e3.13}). \end{defn}
\begin{thm} The Oracle Gen-GBH controls the FDR under Assumptions \ref{assump1} and \ref{assump2}. \end{thm}
This theorem follows from Result \ref{result1} and Lemma \ref{lemma3}.

\subsubsection{Data-Adaptive Procedure} Just like the Gen-GBH, we integrate the DAHeir-GBH into the simultaneous classification setting to propose our data-adaptive version of the Gen-GBH. More specifically, let
\begin {eqnarray} \label{e3.14}\frac{1}{\widehat{W}_i} = \frac{1}{S} \sum_{s=1}^S \frac {1}{\widehat{W}_i(s)}, \end{eqnarray}
where, for each $s$, $\widehat{W}_i(s)$ is defined as in (\ref{e3.6}) and (\ref{e3.7}) using $\hat {\pi}^0_{s;g_1 \cdots g_l}$, for $l=1, \ldots, L_s$, we formally define our proposed data-adaptive procedure for generalized classification of hypotheses as follows:

\begin{defn} [Data-Adaptive Generalized Grouped BH (DA-Gen-GBH)] The level $\alpha$ BH procedure applied to $\widehat{W}_iP_i$, for $i=1, \ldots, N$, where  $\widehat{W}_{i}$ is defined as in (\ref{e3.14}). \end{defn}
\begin{thm}
The DA-Gen GBH controls FDR at level $\alpha$, under the assumption that the p-values are independent.
\end{thm}
The proof of this theorem follows trivially from those of theorems \ref{theorem2} and \ref{theorem4}.
\section{Simulations}\label{SecSimulations}

\begin{figure}
\centering
\includegraphics[width=0.7\textwidth]{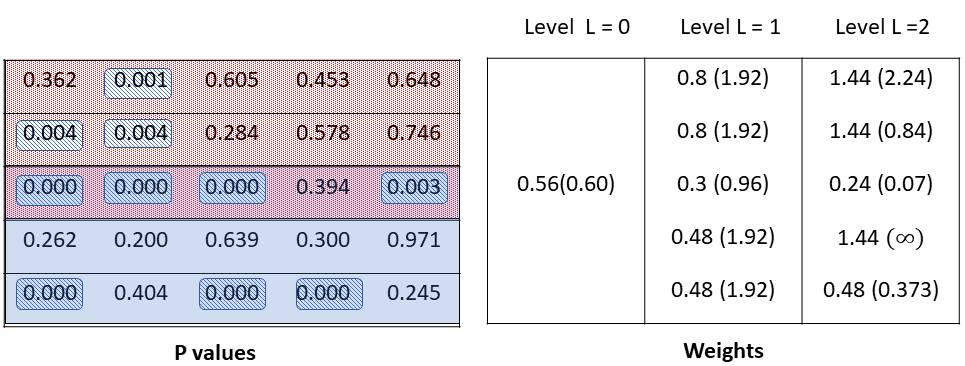}	
\caption{Figure showing the working of the  Heir-GBH and DAHeir-GBH in contrast with the oracle and Adaptive BH. There are $25$ p-values classified in 2 hierarchical levels and the significant p-values are highlighted. The two different shades denote the groups at level $L=1$ and the five rows denote the groups at level $L=2$. The weights assigned by each of the adaptive methods are shown in the adjacent table, and the ideal weights calculated by the oracle methods in each case are mentioned in brackets. At level 0, the Heir-GBH and DAHeir-GBH assign the same weights to the p-values as the BH and Adaptive BH, respectively. }
\label{figsimulation1}\end{figure}

{\bf An example demonstrating the advantage of the Heir-GBH over the BH and respectively, their data-adaptive versions, in a hierarchical classification setup.} Through a short example, we illustrate the superior distinguishing power of the Heir-GBH and the DAHeir-GBH over existing comparable testing procedures, the oracle BH and its data-adaptive version.
We consider the setup similar to that in the first row of Figure \ref{fighierarchical1}, consisting of $N=25$ hypotheses that are hierarchically classified in two levels, with the p-values corresponding to significant hypotheses highlighted. The Adaptive BH disregards classification structure and assigns to each p-value $P_i, i=1, \ldots, N$, the constant estimate of weight (see \cite{Storeyetal2004}):
\begin{align}\label{adaptiveweight}
\widehat{w} = \frac{N-R_N(\lambda) + 1}{N(1-\lambda)}
\end{align}
where $R_N(\lambda) = \sum_{i=1}^N\mathds{1}(P_i \le \lambda)$, with $\lambda$ being set at $0.5$, for the Adaptive BH as well as for the DAHeir-GBH.
In Figure \ref{figsimulation1}, the layout of the p-values is shown, alongwith a table showing the weights assigned by the four methods in three stages.  At level $0$, since there is no grouping, the hierarchical methods (with choice of weights as in (\ref{e3.2}), (\ref{e3.3}), and (\ref{e3.6}), (\ref{e3.7})) are equivalent to the BH in its oracle and data-adaptive forms, respectively. As seen from the table of weights, the Adaptive BH, equivalent to the DAHeir-GBH at $L=0$, assigns constant weight $0.56$, while the oracle BH, equivalent to the Heir-GBH at $L=0$ assigns the weight $0.6$ to all the p-values. At level $L=1$, the weights assigned by the Heir-GBH and DAHeir-GBH are more sensitive to the significant state of the hypotheses than the oracle BH or the Adaptive BH. The five hypotheses at the intersection of the two groups at level one are assigned the smallest weights, appropriately accounting for the effects of both overlapping parent groups. At level $L=2$, the hierarchical procedures are more sensitive to the significant states of the p-values; the highest weights are assigned to the members of group 4 at level two, that does not contain any non-null p-values, and the smallest weights are assigned to the members of group 3 at level two, that are in the intersection of the overlapping groups at level one and contain the largest number of non-null p-values. The weights are sensitive to the effect of grouping on the significance of the individual p-value, and accuracy of the weights increase with the details of the hierarchical structure.

We further investigate the performance of the Heir-GBH in comparison with the oracle BH and the p-filter algorithm. The DAHeir-GBH is compared to the Adaptive BH and the pfilter algorithm with adaptive weights. The comparisons are made, respectively, in terms of (i) control on FDR and (ii) average power at varying densities of signals. We consider $N=5000$ hypotheses in a hierarchical setup of two levels. The layout resembles a section of the hierarchical setup of the EEG dataset we analyzed in the following secion. At the first level, there are two groups, i.e., $m_1=2$, the first group $G(g_1=1)$ comprising of $n_{g_1=1} = 3000$ hypotheses and the second group $G(g_1=2)$ consisting of $n_{g_1=2} =2500$ member hypotheses. The groups overlap and there are $500$ hypotheses at the intersection of $G(g_1=1)$ and $G(g_1=2)$. At level 2, each of these two groups are further classified into smaller groups each containing $100$ members, i.e., $m_2(G(g_1=1)) = 30$, $n_{g_1g_2} = n_{1g_2} = 100, \forall g_2 = 1, \ldots,  m_2(G(g_1=1))$ and $m_2(G(g_1=2)) = 25$, $n_{g_1g_2} = n_{2g_2} = 100, \forall g_2=1,\ldots, m_2(G(g_1=2))$. For the sake of simplification, the hypotheses at the intersection of the two groups at level one are classified into $5$ groups, each of size $100$ at level two.  Thus at level $L=2$, there are $50$ groups, each of size $100$.

The simulation setting was designed using the following steps:
\begin{enumerate}
	
	\item [1.] Generate the following arrays to simulate the impact of classification at each level of grouping. 
	\begin{enumerate}
		\item $\boldsymbol{\Theta}_{L=0}$ as an $m \times n$ random matrix of i.i.d. Ber$(1-\pi_{0})$, where $m=50, n=100$. The elements of the matrix $\Theta_{L=0}$ represent the state of the individual hypotheses at level $L=0$.
		
		\item $\boldsymbol{\Theta}_{L=1}$ as defined below demonstrates the influence of second level of grouping on the set of hypotheses.
		\begin{align*}
		\boldsymbol{\Theta}_{L=1} = 
		\begin{pmatrix}
		\theta_{G(g_1 = 1)}
		\\ \\
		\theta_{G(g_1=1) \cap G(g_1=2)}\\ \\
		\theta_{G(g_1=2)}
		\end{pmatrix}
		\star
		\begin{pmatrix}
		\boldsymbol{1}_{25}\boldsymbol{1}^T_n \\ \\
		\boldsymbol{1}_{5}\boldsymbol{1}^T_n \\ \\
		\boldsymbol{1}_{20} \boldsymbol{1}^T_n \\
		\end{pmatrix}.
		\end{align*}
		$	\theta_{G(g_1=1)}$ and $\theta_{G(g_1=2)}$ are random observations obtained from Ber$(1-\pi_{1})$ to simulate the effect of the groups at level $L=1$. In order to reflect the overlapping significance of $G(g_1=1)$ and $G(g_1=2)$ we draw a sample $\theta_{G(g_1=1)\cap G(g_1=2)} \sim $ Ber$(1-\pi^*_{L=1})$, where $\pi^*_{1} \le \pi_{1}$. Each of these effects are imposed on the respective hypotheses by $\boldsymbol{\Theta}_{L=1}$.
		
		\item $\boldsymbol{\theta}_{L=2}$ as a random vector of $m=50$ i.i.d. Ber$(1-\pi_{2})$, reflects the grouping effect at level $L=2$. 
		
	\end{enumerate}

	\item [2.] Obtain
	\begin{align}\label{e23}
	& \boldsymbol{\Theta} = \boldsymbol{\Theta}_{L=0}\star \boldsymbol{\Theta}_{L=1}\star\boldsymbol{\theta}_{L=2}\boldsymbol{1}^T_n,
	\end{align}
	where  $A\star B$ denotes the Hadamard product between matrices $A$ and $B$, and $\boldsymbol{1}_a$ representing the $a$-dimensional vector of $1$'s. 
	
	\item [3.] Given $\boldsymbol{\Theta}$, generate a random $m \times n$ matrix $\mathbf{X}=((X_{gh}))$ as follows:
	\begin{align}\label{esupp1.2}
	\mathbf{X}  = & \quad \mu\boldsymbol{\Theta} + \sqrt{(1-\rho_{L1})(1-\rho_{L2})}\mathbf{Z}_{mn} + \sqrt{(1-\rho_{L1})\rho_{L2}}\mathbf{Z}_{m}\boldsymbol{1}_n^T + \nonumber \\   & \quad \sqrt{\rho_{L1}(1-\rho_{L2})}\boldsymbol{1}_m\mathbf{Z}_n^T + \sqrt{\rho_{L1}\rho_{L2}}Z_0\boldsymbol{1}_m\boldsymbol{1}_n^T,
	\end{align} having generated $\mathbf{Z}_{mn}$ as $m \times n$ random matrix, $\mathbf{Z}_{m}$ as $m$-dimensional random vector, and $\mathbf{Z}_{n}$ as $n$-dimensional random vector, where $m=50, n=100$, each comprising i.i.d. $N(0,1)$ samples, and $Z_0$ as an additional $N(0,1)$ random variable. The quantities $\rho_{L1}$ and $\rho_{L2}$ measure the strength of correlation among the p-values in each group at levels $L=1$ and $L=2$ respectively. For simplicity, we assume same $\rho_{L1}$ and $\rho_{L2}$ in all groups at respective levels of grouping.
	
	\item [4.] Apply each of the DAHeir-GBH and BH procedures at FDR level $\alpha=0.05$, and the p-filter process with $\alpha = 0.05$ at each level of grouping, for testing $H_{i}: \text{E}(X_{i}) = 0$ against $K_{i}: \text{E}(X_{i}) > 0$, simultaneously for all $i=1, \ldots, N$, in terms of the corresponding weighted p-values. Note the proportions of false rejections among all rejections and correct rejections among all false nulls.
	
	\item [5.] Repeat Steps 1-4 500 times to simulate the values of FDR and power for each procedure by averaging out the corresponding proportions obtained in Step 4.\
	
\end{enumerate}

\begin {remark} \rm Note that
\begin{eqnarray*} \label{es1}
	\mathbf{vec}({\mathbf{X}}) \sim \text{N}_{m n}(\mathbf{vec}(\mu \boldsymbol{\Theta}), \Sigma_{L2} \otimes \Sigma_{L1}),
\end{eqnarray*}
where $\Sigma_{L1} = (1 -\rho_{L1})I_n + \rho_{L1}\boldsymbol{1}_n\boldsymbol{1}_n^T, \; \rho_{L1} \in [0, 1)$, and $\Sigma_{L2} = (1 -\rho_{L2})I_m + \rho_{L2}\boldsymbol{1}_m\boldsymbol{1}_m^T, \; \rho_{L2} \in [0,1).$ Thus, the test statistics are allowed to have different types of dependence structure by appropriately setting the value of $\rho_{L1}$ and/or $\rho_{L2}$ at $0$ or some non-zero value.

The formulation in equation (\ref{e23}) allows us to determine the significant state of each hypothesis, subject to the state of significance of its parent groups at all levels. This helps to demonstrate the true effect of the underlying hierarchical structure in the simulation studies. Since the groups at level $L=1$ overlap, we can regulate the density of signals in intersecting region using the following
\begin{eqnarray}\label{esupp1.3}
\pi & = & 1 - (1 - \pi_0)(1-\pi^*_{1})(1 - \pi_{2}),
\end{eqnarray} 
and in the non-intersecting regions by the following
\begin{eqnarray}\label{esupp1.4}
\pi & = & 1 - (1 - \pi_0)(1-\pi_{1})(1 - \pi_{2}),
\end{eqnarray} 

Here $\pi$ represents the proportion of true nulls in the entire set of $mn$ hypotheses in terms of the proportions of groups at level 2 i.e., ($1-\pi_{2}$), and at level $1$, i.e, ($1-\pi_{1}$) (or $(1-\pi^*_{1})$ in the overlapping region) containing signals, and the proportion of individual hypotheses that are significant $(1-\pi_0)$, when no classification is imposed, i.e., at level $L=0$.  \end {remark}

We set $\mu=0$ for true null hypotheses and $=3$ for true signals. We also set $\pi_{2} = 0.5$, $\pi_{L=1} =0.5$, $\pi^*_{L=1} =0.25$ and increase $1-\pi_0$ from $0$ through $1$ in order to simulate different densities of signals using  (\ref{esupp1.3}) and (\ref{esupp1.4}).

\begin{figure}[H]
	\centering
	\includegraphics[width =0.7\textwidth]{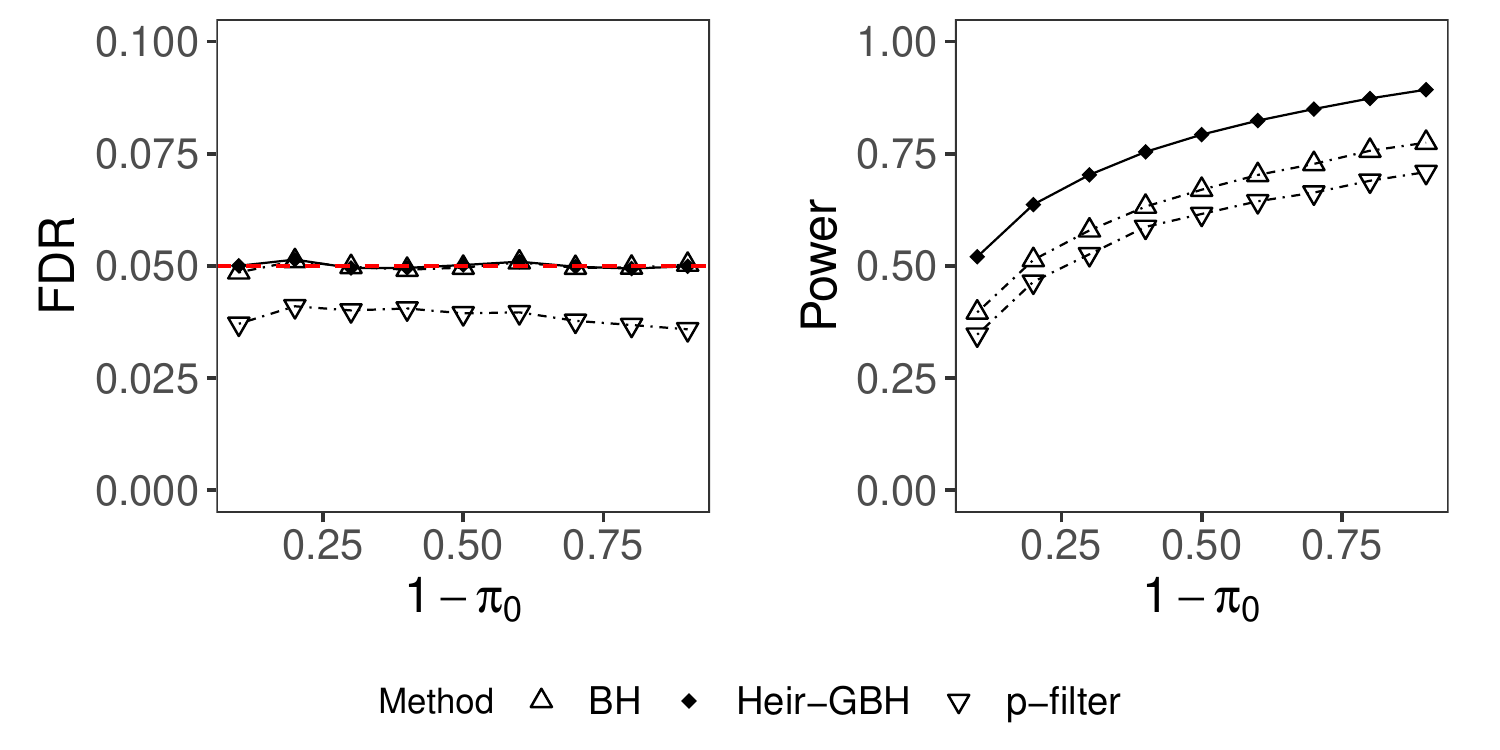}	
	\caption{Comparison of the Heir-GBH with the BH and the p-filter algorithm under independence.}
	\label{figS1}
\end{figure}

\begin{figure}[H]
	\centering
	\includegraphics[width =0.7\textwidth]{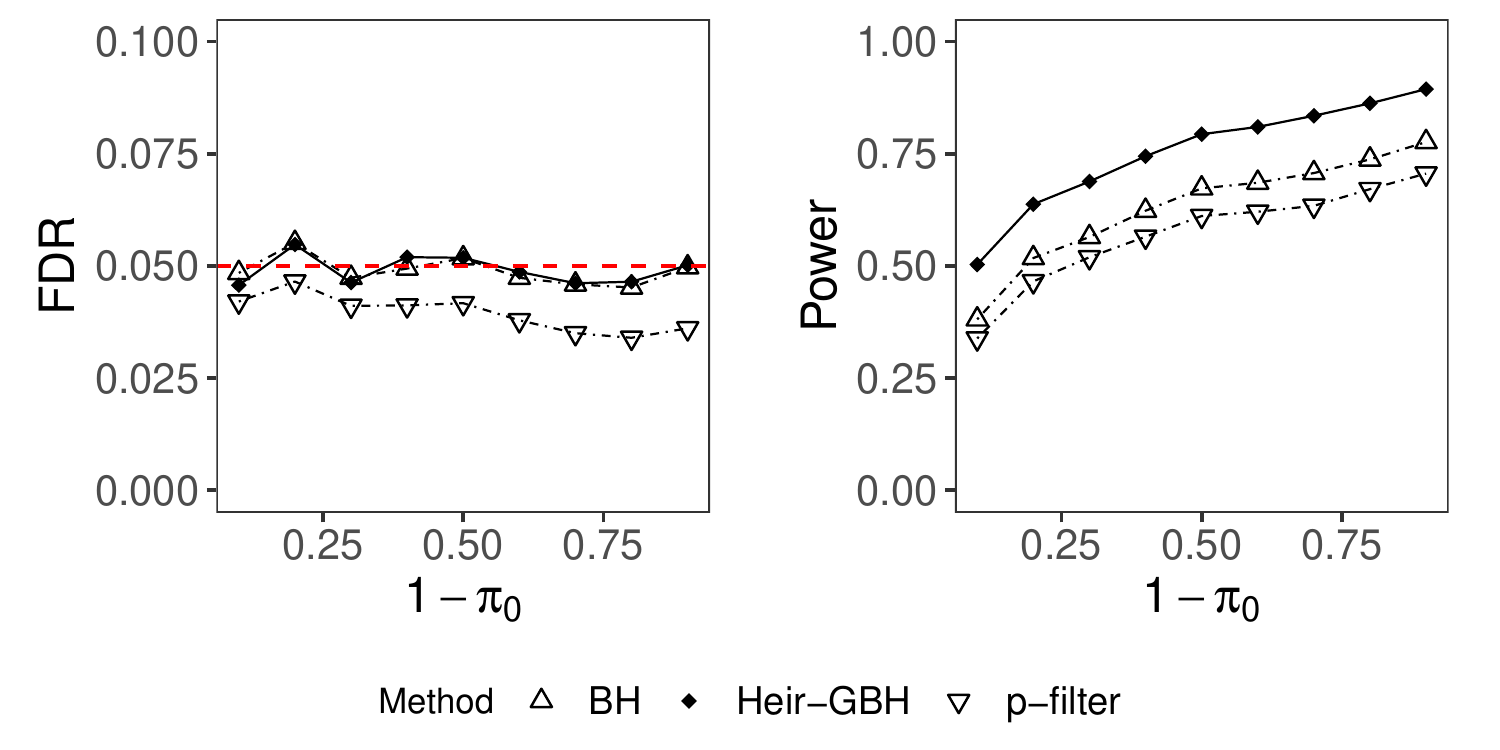}	
	\caption{Comparison of the Heir-GBH with the BH and the p-filter algorithm  under the assumption that the p-values are dependent.}
	\label{figS2}
\end{figure}

\begin{figure}[H]
	\centering
	\includegraphics[width =0.7\textwidth]{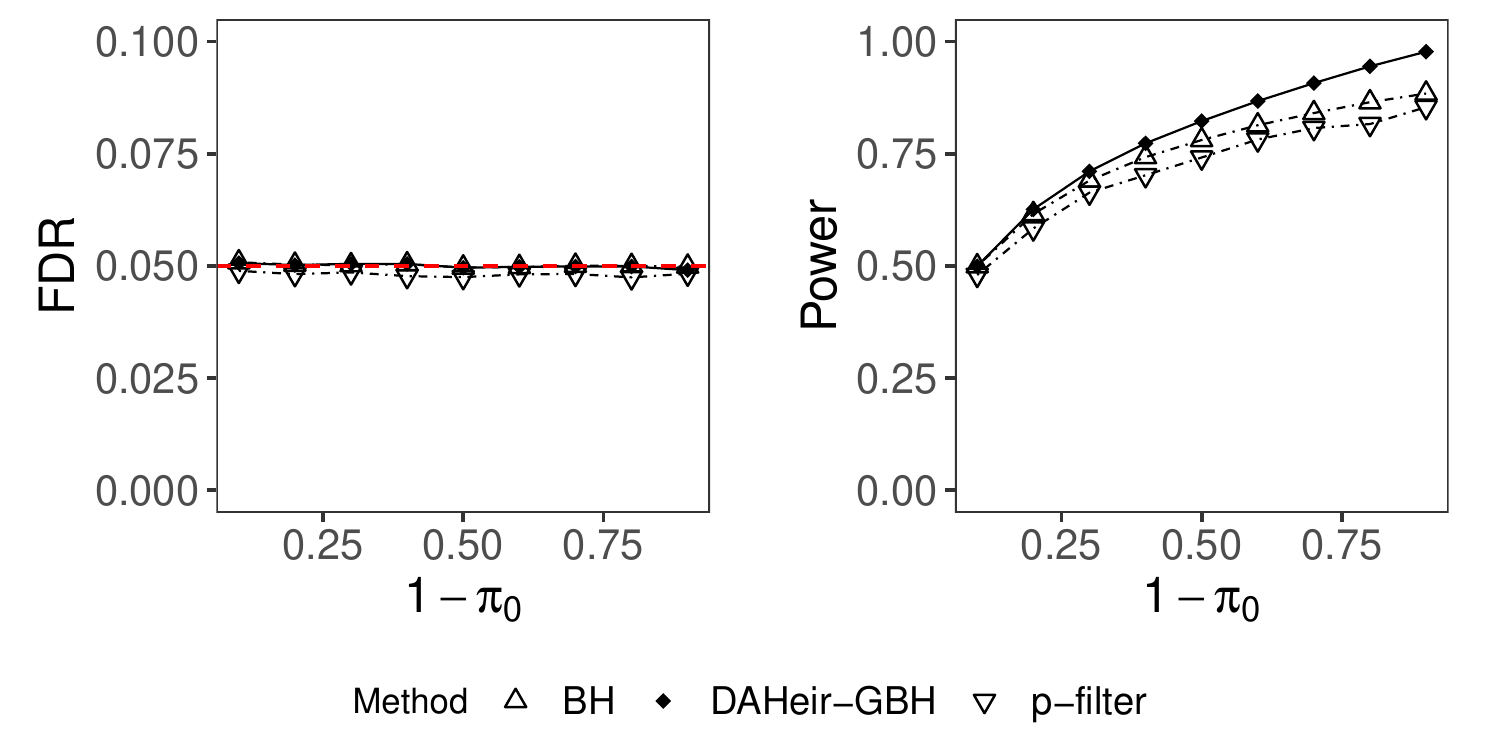}	
	\caption{Comparison of the DAHeir-GBH with the Adaptive BH and the p-filter algorithm when the p-values are independent.}
	\label{figS3}
\end{figure}

\noindent
\textbf{Comparison of the Heir-GBH with the Oracle BH and the p-filter algorithm.}
Figure \ref{figS1} compares the performance of the Heir-GBH (as defined in Definition 2) and the oracle BH  and the p-filter process under the assumption that the p-values are independent.
Figure \ref{figS2} shows the comparison under the assumption that the p-values are positively dependent. For this purpose, we set $\rho_{L1} = 0.3$ and $\rho_{L2}=0.4$ in (\ref{esupp1.2}). 
In either case, the comparisons show that the pfilter algorithm is more conservative than the BH and the Heir-GBH in terms of controlling the FDR, and incorporating hierarchical weights makes the Heir-GBH significantly more powerful than the other two methods at all levels of densities of signals.
\noindent
\textbf{Comparison of the DAHeir-GBH with the Adaptive BH and the p-filter algorithm.}
In order to compare with the DAHeir-GBH (as in Definition 3), we choose the set of weights proposed by \cite{Storeyetal2004} for the Adaptive BH, which coincides with the weight used by DAHeir-GBH at level $L=0$. That is, each p-value is assigned the estimated weight $$\widehat{w} = \frac{N-R_N + 1}{N(1-\lambda)},$$ where $R_N = \sum_{i=1}^N \mathds{1}(P_i \le \lambda)$. Under the assumption that the p-values are independent, all three methods control FDR, but the DAHeir-GBH is more powerful than the others, as shown in figure \ref{figS3}. Besides the simulation settings used for the comparison of the oracle counterparts, we set $\lambda=0.5$ for the DAHeir-GBH and adaptive BH procedures. For the pfilter algorithm, we set $\lambda^{(m)} = \lambda$ for all $m = 0,1, 2$(see section 2 of \cite{ramdas2019}). 

\begin{figure}[H]
	\centering
	\includegraphics[width =0.7\textwidth]{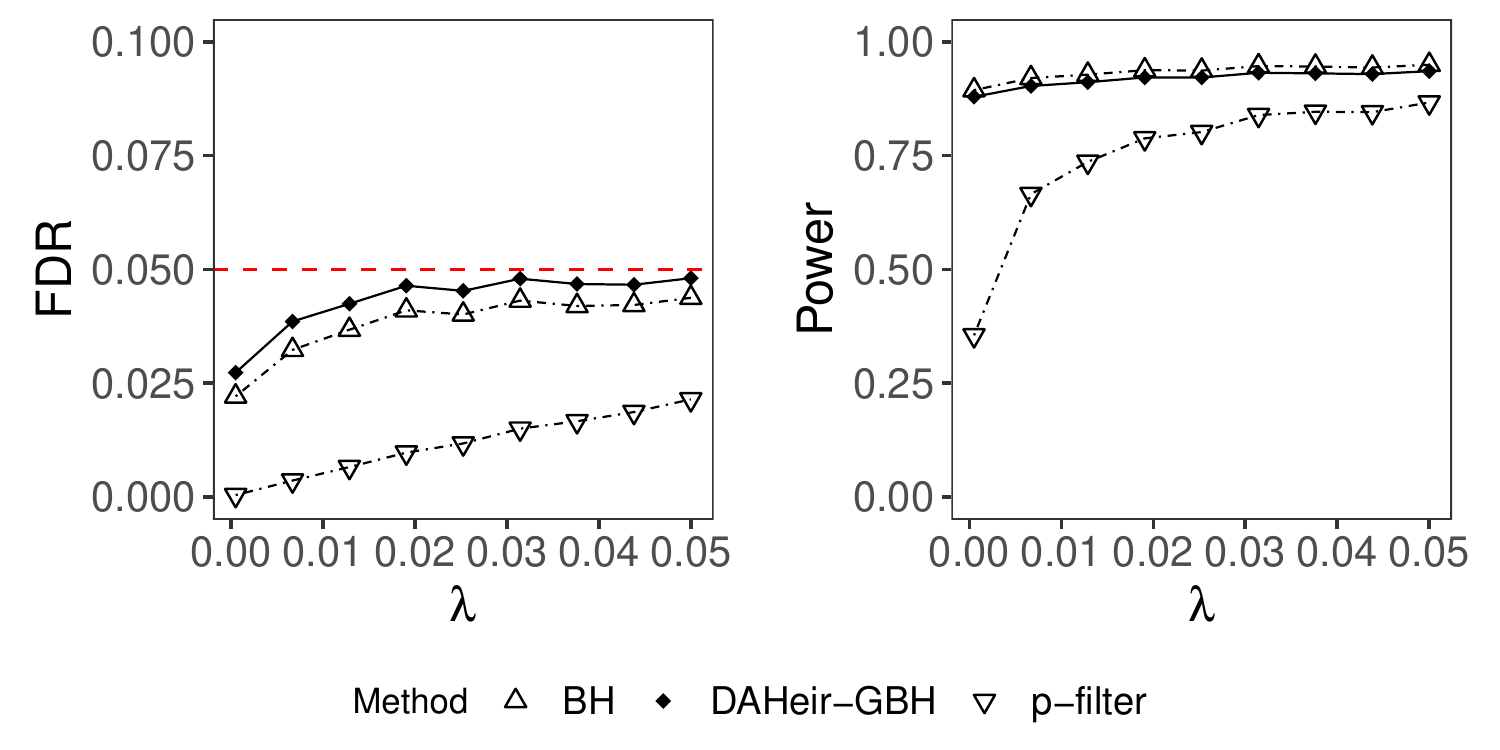}	
	\caption{Comparison of the DAHeir-GBH with the adaptive BH and the p-filter algorithm under the assumption that the p-values satisfy the PRDS condition.}
	\label{figS4}
\end{figure}

When the p-values satisfy the PRDS condition, the DAHeir-GBH still controls FDR under certain conditions. We set $\pi_{1} =\pi^*_{1}=\pi_{2}=0$ so that the signals are uniformly distributed in all groups at levels one and two, and the variability comes from $\pi_0 = 0.3$. We choose $\lambda < \alpha=0.05$ for the DAHeir-GBH and the adaptive BH procedures. For the p-filter algorithm, we set $\lambda^{(m)}  =\lambda$, for $m = 0,1,2$. The within group dependence at levels one and two are characterized by $\rho_{L1} = 0.3$ and $\rho_{L2} = 0.4$ respectively. The comparative performances of the three procedures are shown in figure \ref{figS4}. All procedures conservatively control the FDR at all levels of $\lambda$, and the proposed DAHeir-GBH has comparable power with the adaptive BH. 

Interested readers are  referred to \cite{NANDI2021} for numerical results on the simultaneous classification setup. 

The simulation results here and in \cite{NANDI2021} assert that our proposed methods in the hierarchically as well as simultaneously grouped setting are able to control FDR, at least under independence, and are more powerful than existing procedures. Since the Gen-GBH and DAGen-GBH combine the hierarchical and simultaneous procedures, their performances would evidently be better than existing competing methods.

\section{Analysis of the EEG patterns through DAGen-GBH}
\begin{figure}[H]
$\begin{array}{rcl}
\begin{subfigure}[b]{0.5\textwidth}
\includegraphics[width=\textwidth]{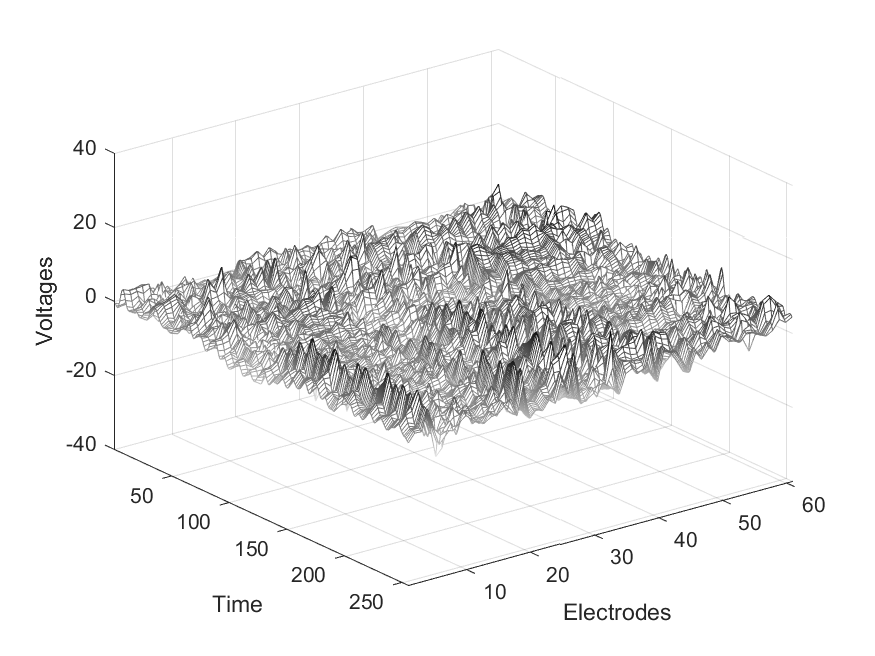}
\caption{Left Hemisphere}
\label{control}
\end{subfigure} &
\begin{subfigure}[b]{0.5\textwidth}
\includegraphics[width=\textwidth]{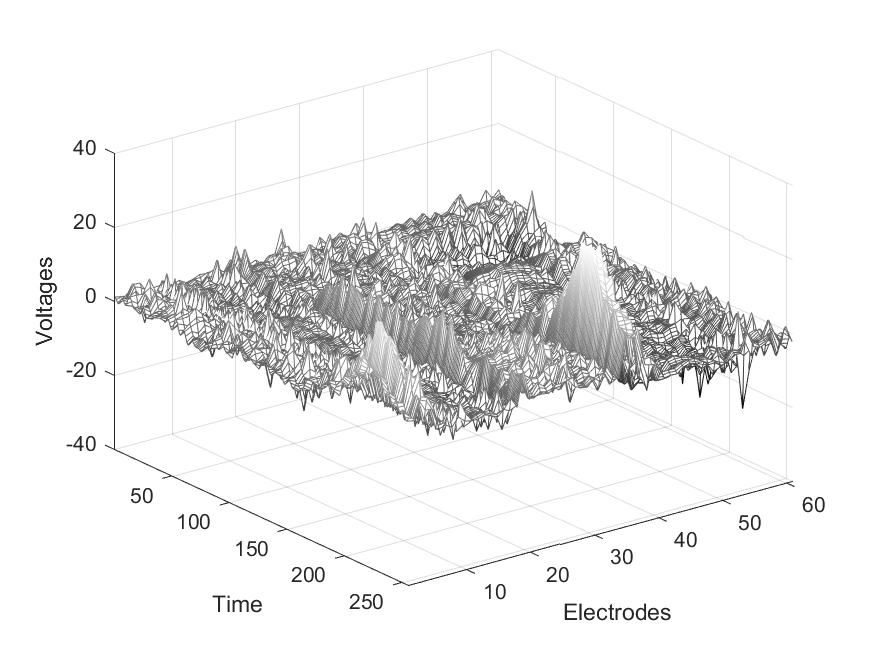}
\caption{Right Hemisphere}
\label{alcoholic}
\end{subfigure}
\end{array}$
\caption{EEG measurements recorded from two subjects, one each in the control and alcoholic groups.}
\label{fig1}
\end{figure}

We apply our DAGen-GBH procedure to analyze the data in \url{http://kdd.ics.uci.edu/databases/eeg/eeg.data.html}, which arises from a large study examining  EEG correlates of genetic predisposition to alcoholism. The experiment was conducted on a control group and a treatment group comprising of alcoholic subjects. 
For our analysis, we consider $10$ subjects each in the control group and alcoholic group. On each subject, ten trials were performed. During each trial, a picture from \cite{snodgrassandvanderwart} was presented to the subject, while EEG activity in the form of voltage fluctuations (in microvolts) were recorded at $256$ time points (each time point being $1/256$th of a second) from $61$ electrodes placed on the subject's scalp.
Figure~\ref{fig1} shows an example of the EEG pattern, averaged over measurements obtained from ten trials, for two subjects, one each from the control and alcoholic groups. The $x$ and $y$ axes correspond to time and scalp electrodes, respectively, and the $z$-axis corresponds to the EEG measurements. It illustrates a clear difference in the voltage fluctuation patterns for the two subjects over time and electrodes, which leads us to the following driving question underlying this study: 

\begin {quote}
\renewcommand\baselinestretch{1}\tiny\normalsize
\noindent  {\it At which pairs of electrodes, are the EEG measurements significantly different between the two groups of subjects?}
\end{quote}
Answering this question can shed further insight into brain dysfunction and regions impacted by alcoholism. 
In the literature, case-control studies on EEG measurements, such as this, has often been used to diagnose brain dysfunction and evaluate the performance of relevant treatment procedures (for examples, see \cite{Centorrino2002}, \cite{Oscar-Berman2007},  etc.). 
Specifically, we consider simultaneous testing of the following null hypotheses.
\begin{eqnarray}\label{e19}
	& H_{ghk}:& \mbox{Between the alcoholic and control groups, there is no difference in the mean  } \\ & &\text{EEG measurement recorded at any time point at a pair of elcetrodes } (g,h), \nonumber
\end{eqnarray} for $g = 1,\ldots, 61$, $h = 1, \ldots, 61$.
We have $95,2576$ such hypotheses.
The hypotheses in~(\ref{e19}) are laid out in a generalized classification setup as described in Section \ref{section3.3}.
The placement of electrodes and anatomy of the brain provide valuable information about the spatial arrangement of the hypotheses. 
All electrode positions are described by the \textit{International 10-20} system, which is based on the location of an electrode and the underlying area of the cerebral cortex. Subsequently, an electrode can be classified according to the region of the brain it is reading from. Letter codes are used to describe the electrode positions in the six regions of the brain, i.e., pre-frontal (Fp), frontal (F), central (C), parietal (P), occipital (O) and temporal (T).
The positions of some electrodes are not related to particular regions of the brain. These are placed on intermediate positions halfway between the electrodes placed according to the 10-20 system. These are named using two letter codes, for example, the electrodes on the margin of frontal and temporal regions are denoted by FT, those between temporal and parietal regions are denoted by TP, etc.  It is logical to expect that these electrodes reflect the neuronal activities of brain regions between which they lie, which lead us to consider classification of such electrodes as not mutually exclusive, and these are considered to be members of either regions. Each hypothesis is associated to a pair of electrodes, one corresponding to the control and the other corresponding to the alcoholic group.
 In either group, at level $L=0$, without any classification, there is a single group of $N=952576$ hypotheses. At level $L=1$, these hypotheses are clustered according to the six brain regions. In the ultimate level $L=2$, a hypothesis is classified by its corresponding electrode. There are $61$ such groups in level $2$, each consisting of $256$ member hypotheses. 
 Two brain regions are considered to be overlapping if there is an electrode placed at their margins, with this electrode representing a common group of hypotheses. Each brain region overlaps with at least one and at most two other brain regions.

The classifications mandated by the control and treatment groups are simultaneously imposed on the set of $N$ hypotheses. Consequently, there are two simultaneously imposed hierarchical classification structures, according to electrodes and brain regions on the hypotheses.

Prior to computing the test statistics, the time series observations are detrended and cyclic/phase patterns are removed. 
The p-values corresponding to the hypotheses in (\ref{e19}) are obtained from 
two-sample {$t$-test} statistics calculated from the de-trended sets of observations.

We apply the DAGen-GBH, with the choice of weights derived from (\ref{e3.14}), and attach the following weight to the p-value $P_i$, for each $i=1, \ldots, N$:
\begin{align}\label{e4.1}
&\widehat{W}_{i} = \left[\frac{1}{2} \left(\frac{1}{\widehat{W}_{i}(s=1)} + \frac{1}{\widehat{W}_{i}(s=2)}\right)\right]^{-1},
\end{align}
where $s=1$ refers to the classification due to alcoholic group, and $s=2$ refers to the classification of the p-values due to the control group.
From (\ref{e3.6}), we get 
\begin{align*}
 \frac{1}{\widehat{W}_{i}(s=1)} =  \left(\frac{1}{N}\sum_{g_1=1}^{m_1}\sum_{g_2=1}^{m_2} \frac{(n_{g_1g_2} -R_{g_1g_2}+1)/(1-\lambda)}{\widehat{w}_{g_1g_2}}\right)^{-1} \times 
  \sum_{g_1}\sum_{g_2}\frac{1}{\widehat{w}_{g_1g_2}}\mathds{1}(i \in I_{g_1g_2}),
\end{align*}
for all $i$, where $\widehat{w}_{g_1g_2} = \hat{\pi}^0\cdot \frac{\hat{n}^0_{g_1g_2}}{\hat{n}^0_{g_1}}\cdot m_2,$ which is given by (\ref{e3.7}).
Here $N=952516$, $n_{g_1}$ is the size of the $g_1$th brain region, $m_1 = 6$, the number of brain regions and $m_2 = m_2(G(s=1; g_1))$ is the number of electrodes in the $g_1$th brain region, each electrode containing $n_{g_1g_2}=15616$ member p-values.
$\widehat{W}_i(s=2)$ is similarly defined for the control group.

\begin{figure}
\centering	
\includegraphics[width=0.4\linewidth]{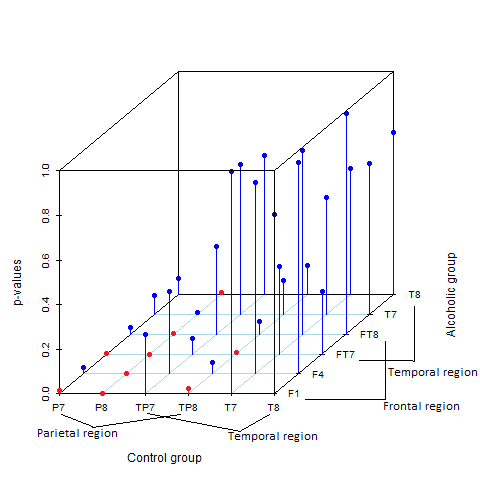}
\caption{A section of the classification structure of the p-values. Six electrodes and their corresponding brain regions are shown from the alcoholic and control groups. These brain regions overlap over two electrodes in each case. The smaller highlighted p-values represent signals identified by the DAGen-GBH, these are used merely to present the working idea of the procedure.
}	
\label{fig3}
\end{figure}

Using $\lambda=0.5$ and $\alpha = 0.05$ we obtain $31914$ rejections. 
In comparison, the Adaptive BH, with the choice of weights as in (\ref{adaptiveweight}), yields a set of $28286$ discoveries, approximately $98\%$ of which is a subset of those made by the DAGen-GBH.
Figure~\ref{fig3} graphically represents a section of the classification structure of the p-values and visualizes the signals identified by the DAGen-GBH.

\begin{figure}[!h]
 	
	\begin{floatrow}
		\ffigbox{\includegraphics[scale = 0.75]{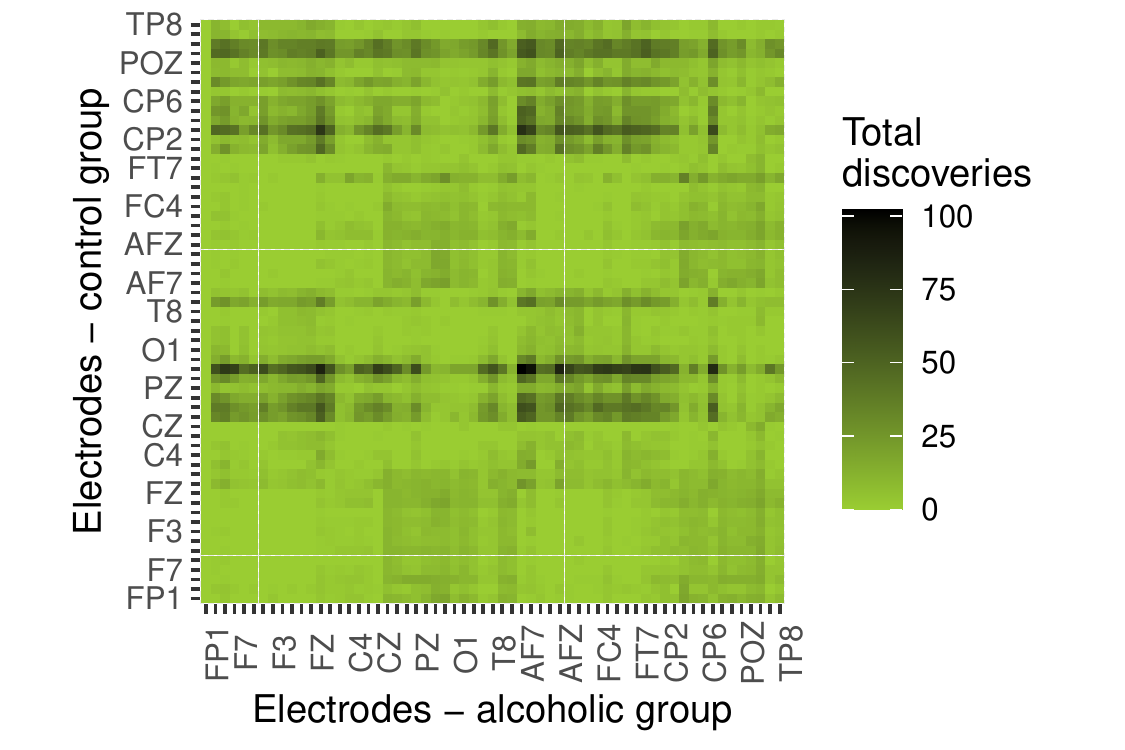}}{\caption{Signals discovered at each combination of electrode pairs}\label{fig5}}
		\ffigbox{\includegraphics[scale = 0.75]{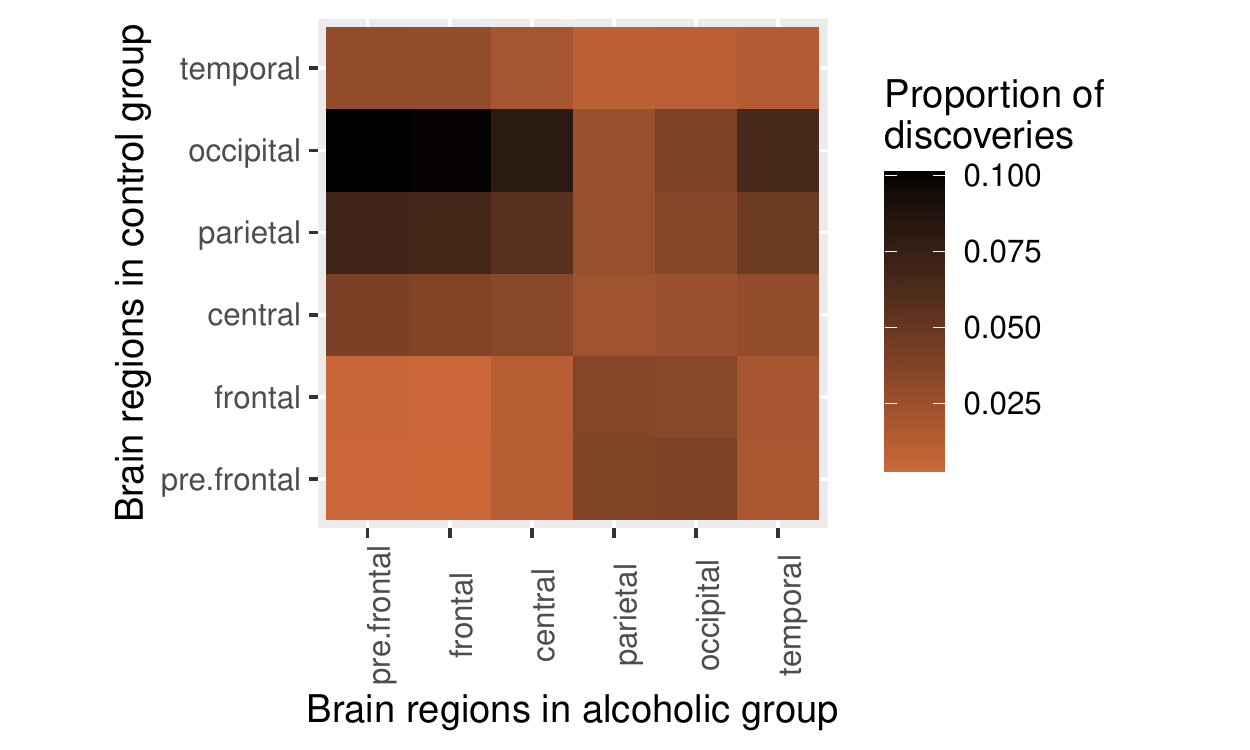}}{\caption{Proportion of signals discovered at each pair of brain regions, one from the alcoholic and the other from the control group.}\label{fig6}}
	\end{floatrow}
\end{figure}

It is interesting to see the distribution of discoveries made at each pair of electrodes, and each pair of brain regions. Figure~\ref{fig5} shows the number of discoveries made at each pair of electrodes from the control and alcoholic groups.  Figure~\ref{fig6} shows the proportion of discoveries made at each pair of brain regions from the control and alcoholic groups. The largest proportion of differences in EEG signals are noted between the pre-frontal and frontal regions in the alcoholic groups and parietal and occipital regions in the control group. Further insights into the discoveries would entail expert knowledge on such studies of neurological disorders. 


\section{Conclusion}


Multiple testing has a long history of being applied to a variety of neuroimaging data (\cite{nichols2003}, \cite{Helleretal2006}, \cite{2015arXiv151203397F}, etc), and has been able to successfully draw valid inference from such complex structures, notwithstanding under-utilization of structural information in many cases. BH-type multiple testing procedures are capable to control FDR and identify true signals when the hypotheses are positively dependent. Additionally, tapping structural information in such procedures, though necessary to analyze datasets complex as neuroimaging data, is clearly not straightforward as shown by this article. The classification of hypotheses in the EEG dataset considered here is an assortment of different types of classifications, that need to be systematically studied to extract the effects of grouping on the individual hypotheses. We apply our idea of generalized  classification that combines hierarchical and simultaneous groupings, and permits overlapping groups, to obtain interesting results pertaining to abnormal effect of alcohol on the functioning of the different regions of the brain, as captured by EEG signals.

This article makes significant contribution to show that there is substantial scope of modernizing the current framework of multiple hypotheses testing to address complex structures of hypotheses. We have broadened the idea of using weighted BH type procedure to address more general types of classification structure of hypotheses, hierarchical and/or simultaneous, and accommodate overlapping groups which is a plausible situation in a hierarchical setup. Our proposed scheme of weighting p-values in weighted BH type procedure utilizes the structural information from the classification setup to define the weights, that provably control the FDR when the p-values involved satisfy the PRDS condition. The corresponding data-adaptive methods estimate the weights from the data, and 
also control FDR under independence of the p-values involved. Our simulation studies highlight the fact that utilizing elaborate grouping information indeed enhances the power of the weighted BH procedures, over existing similar testing procedures. The simulated scenarios also show that given certain conditions of high density of signals and appropriate choices of the tuning parameter, the proposed data-adaptive procedures also control FDR and are satisfactorily powerful when the p-values involved are positively dependent.

There can be further possibilities to improve upon our idea of weighted multiple testing procedure, by incorporating additional structural information, or extraneous knowledge to devise new weights. Though we successfully applied our method to analyze a very complex dataset to draw interesting and valid inference, its scope extends beyond and can be adapted to datasets with similar or increased structural complexities.

\bigskip
\begin{center}
	{\large\bf Acknowledgements}
\end{center}
We are thankful to Edo Airoldi and Debashis Ghosh for their valuable comments, which helped us shape this paper, and to Aaditya Ramdas for very helpful discussion on the p-filter algorithm.

\newpage
\begin{appendix}
	\begin{center}
		{\large\bf Appendix}
	\end{center}
\end{appendix}
\appendix

\section{Proofs }
\subsection{Proof of Result \ref{result2}} The result follows by noting that
\begin{eqnarray} & & \sum_{i=1}^N \mathds{1} (i \in I^0)\frac{1}{W_i} =  \left (\frac {1}{N} \sum_{g=1}^m \frac{n_g \pi_g^0}{w_g} \right )^{-1} \sum_{i=1}^N \mathds{1} (i \in I^0) \sum_{g':I_{g'} \ni i} \frac{1}{w_{g'}} \nonumber
\\ & = & N \left (\sum_{g=1}^m \frac{n_g \pi_g^0}{w_g} \right )^{-1} \sum_{g'=1}^m \frac{1}{w_{g'}} \sum_{i \in I_{g'}} \mathds{1} ( i \in I^0) = N \left (\sum_{g=1}^m \frac{n_g \pi_g^0}{w_g} \right )^{-1} \sum_{g'=1}^m
\frac{n_{g'} \pi_{g'}^0}{w_{g'}} \nonumber \\ & = & N.  \nonumber \\ \end {eqnarray}

\subsection{Proof of Lemma \ref{lemma1}} The lemma follows from Result \ref{result2} and 
the following:

$\bullet$ In the hierarchical setup with overlapping groups, Condition 1 is given as follows
\begin {eqnarray}\label{e9.1}
\sum_{i \in I_0} \frac{1}{W_i} \sum_{g_1=1}^{m_1} \cdots \sum_{g_L=1}^{m_L} \mathds{1}  \left (i \in I_{g_1 \cdots g_L} \right ) = N
\end {eqnarray}

$\bullet$ With groups not overlapping, we have $W_i= w_{g_1 \cdots g_L}$, for each $P_i \in G(g_1 \cdots g_L)$. In such a case too, $W_i$ satisfies Condition 1. This can be seen by substituting the expression of $w_{g_1 \cdots g_L}$ from (\ref{e3.4}), to the left-hand side
of (\ref{e9.1}) which in this case equals to
\begin {eqnarray} & & \sum_{g_1=1}^{m_1} \cdots \sum_{g_L=1}^{m_L} \sum_{i \in I_{g_1 \cdots g_L}^0}\frac{1}{w_{g_1 \cdots g_{L}}} \nonumber \\ & = & \sum_{g_1=1}^{m_1} \cdots \sum_{g_L=1}^{m_L} \sum_{i \in I_{g_1 \cdots g_L}^0}\frac{1}{w_{g_1 \cdots g_{L-2}}}\frac{\pi_{g_1 \cdots g_{L-1}}^0}{1- \pi_{g_1 \cdots g_{L-1}}^0} \frac{1-\pi_{g_1 \cdots
		g_{L}}^0}{\pi_{g_1 \cdots g_{L}}^0} \nonumber \\ & = & \sum_{g_1=1}^{m_1} \cdots \sum_{g_{L-2}=1}^{m_{L-2}}\sum_{i \in I_{g_1 \cdots g_{L-2}}^0} \frac{1}{w_{g_1 \cdots g_{L-2}}}, \nonumber \end{eqnarray} which ultimately
equals $\sum_{g_1=1}^{m_1}n_{g_1} {\pi}_{g_1}^0 /{\pi}^0$ or $\sum_{g_1=1}^{m_1}n_{g_1} (1-{\pi}_{g_1}^0)/(1-{\pi}^0)$, depending on whether $L$ is even or odd, each of which equals $N$.
\subsection{Proof of Theorem 2}
The proof of the theorem follows from Result \ref{result3}, using the following arguments.

First we need to prove $\hat{w}_{g_1\ldots g_L}$ as defined in (\ref{e3.7}), is increasing in each $P_i \in \cup_{g_1=1}^{m_1} \cdots \cup_{g_L}^{m_L} I_{g_1 \cdots g_L} $. In group $G(g_1 \ldots g_L)$, $R_{g_1 \ldots g_L}$ is decreasing in each $P_i \in G(g_1 \ldots g_L)$. Subsequently, 
\begin{itemize}
	\item $\hat{n}^0_{g_1 \ldots g_L}  =  \frac{n_{g_1 \ldots g_L} - R_{g_1 \ldots g_L} +1}{1-\lambda}$ is non-decreasing in each $P_i$,
	\item $\hat{n}^0_{g_1 \ldots g_{l-1}} = m_l \cdot  \hat{n}^0_{g_1 \ldots g_l}$ is also non-decreasing in $P_i$ for all $1 \le l \le L$.
\end{itemize}
Hence, by induction, $\hat{w}_{g_1 \ldots g_L}$ as defined in (\ref{e3.7}), is increasing in $P_i$.
Note that the term
$$  \left ( \frac{1}{N} \sum_{g_1=1}^{m_1} \cdots \sum_{g_L=1}^{m_L} \frac{\hat{n}_{g_1 \cdots g_L}^0}{\hat{w}_{g_1 \cdots g_L} }\right ) \text{ is independent of the p-values, }$$
due to definitions in (\ref{eqn7}), (\ref{eqn8}) and (\ref{e3.7}).
Hence the weight $\widehat{W}_i(\mathbf{P})$, as defined in (\ref{e3.6}) is non-decreasing in each $P_i, i = 1, \ldots, N$.
For each $i \in I^0$, if $R_{g_1\ldots g_L}(\mathbf{P}^{-i},0) = R_{g_1\ldots, g_L} +1$, then 
\begin{align*}
\hat{n}^0_{g_1\ldots g_L}(\mathbf{P}^{-i},0) = \tilde{n}^0_{g_1 \ldots, g_L} = \frac{n_{g_1 \ldots g_L} -  R_{g_1 \ldots g_L}}{1 - \lambda}
\end{align*}
and for any $l \ge 1$, let us denote $\hat{n}^0_{g_1 \ldots g_l}(\mathbf{P}^{-i},0)$ in terms of $\hat{n}^0_{g_1\ldots g_{l-1}}(\mathbf{P}^{-i},0)$, by $\tilde{n}^0_{g_1 \ldots g_l}$, and let us denote $\hat{w}_{g_1 \ldots g_l}(\mathbf{P}^{-i},0)$ in terms of $\tilde{n}^0_{g_1 \ldots g_l}$ by $\tilde{w}_{g_1 \ldots g_l}$.\\	 
In order to prove ${E\left[\sum_{i \in I^0} \frac{1}{\widehat{W}_i(\mathbf{P}^{-i}, 0)}\right] \le N, }$\\
For $L=1$, 
\begin{align}\label{e7.3}
& \;\quad E\left[\sum_{g_1=1}^{m_1}\sum_{i \in I^0_{g_1}}\left(\left(\frac{1}{N}\sum_{g_1 = 1}^{m_1}\frac{\hat{n}^0_{g_1}(\mathbf{P}^{-i},0)}{\hat{w}_{g_1}(\mathbf{P}^{-i},0)}\right)^{-1} \cdot \frac{1-\lambda}{n_{g_1} - R_{g_1}(\mathbf{P}^{-i},0) +1} \cdot \frac{N}{m_L}\cdot \mathds{1}(I_{g_1} \ni i)\right)\right] \nonumber \\
& =  E\left[\frac{N}{m_L} \cdot \sum_{g_1=1}^{m_1} \left(\frac{n^0_{g_1}-V_{g_1}(\lambda)}{n_{g_1} - R_{g_1}}\right)\right] \le E\left[\frac{N}{m_L} \cdot \sum_{g_1=1}^{m_1} 1\right]  = N
\end{align}

where $V_{g_1}(\lambda)$ is the number of false rejections in the $g_1$th group. For $L\ge 2$, first note that
\begin{align}\label{e7.4}
& E\left[\sum_{i \in I^0} \frac{1}{\widehat{W}_i(\mathbf{P}^{-i}, 0)}\right] = 		
E\left[\sum_{i \in I^0} \sum_{g_1} \ldots \sum_{g_L}\left\{  \left(\frac{1}{N} \sum_{g_1} \cdots\sum_{g_L} \frac{\tilde{n}^0_{g_1 \ldots g_L}}{\tilde{w}_{g_1 \ldots g_L}} \right)^{-1}\frac{\mathds{1}(I_{g_1\ldots g_L} \ni i)}{\tilde{w}_{g_1 \ldots g_L}}\right\}\right] 
\end{align}
Note that, using the formula for $\tilde{w}_{g_1 \ldots g_L} = \tilde{w}_{g_1\ldots g_{L-2}} \cdot \frac{\tilde{n}^0_{g_1 \ldots g_{L}}}{\tilde{n}^0_{g_1 \ldots g_{L-1}}}$ from (\ref{e3.7}),
\begin{align*}
\sum_{g_1} \cdots\sum_{g_L} \frac{\tilde{n}^0_{g_1 \ldots g_L}}{\tilde{w}_{g_1 \ldots g_L}} 
=   \sum_{g_1} \cdots \sum_{g_{L-1}}\frac{\tilde{n}^0_{g_1 \ldots g_{L-1}}}{\tilde{w}_{g_1 \ldots g_{L-2}}}\cdot \sum_{g_L}\frac{1}{m_L}	
=  \sum_{g_1} \cdots \sum_{g_{L-1}}\frac{\tilde{n}^0_{g_1 \ldots g_{L-1}}}{\tilde{w}_{g_1 \ldots g_{L-2}}}
\end{align*}
Hence (\ref{e7.4}) implies $E\left[\sum_{i \in I^0} \frac{1}{\widehat{W}_i(\mathbf{P}^{-i}, 0)}\right] $
\begin{align}\label{e7.5}
& = E\left[\sum_{i \in I^0} \sum_{g_1} \ldots \sum_{g_L}\left\{  \left(\frac{1}{N} \sum_{g_1} \cdots\sum_{g_{L-1}} \frac{\tilde{n}^0_{g_1 \ldots g_{L-1}}}{\tilde{w}_{g_1 \ldots g_{L-2}}} \right)^{-1} \cdot \frac{\tilde{n}^0_{g_1 \ldots g_{L-1}}}{\tilde{w}_{g_1 \ldots g_{L-2}}} \cdot \frac{1 - \lambda}{n_{g_1 \ldots g_L} - R_{g_1 \ldots g_L}} \cdot \frac{\mathds{1}(I_{g_1\ldots g_L} \ni i)}{m_L}\right\}\right]  \nonumber \\
& =   E\left[ \left(\frac{1}{N} \sum_{g_1} \cdots\sum_{g_{L-1}} \frac{\tilde{n}^0_{g_1 \ldots g_{L-1}}}{\tilde{w}_{g_1 \ldots g_{L-2}}} \right)^{-1} \cdot  \sum_{g_1} \ldots \sum_{g_L} \left\{ \frac{n^0_{g_1 \ldots g_L} - V_{g_1 \ldots g_L}}{n_{g_1 \ldots g_L} - R_{g_1 \ldots g_L}} \cdot \frac{1}{m_L} \cdot  \frac{\tilde{n}^0_{g_1 \ldots g_{L-1}}}{\tilde{w}_{g_1 \ldots g_{L-2}}}\right\}   \right]    \nonumber \\
& \le E\left[ \left(\frac{1}{N} \sum_{g_1} \cdots\sum_{g_{L-1}} \frac{\tilde{n}^0_{g_1 \ldots g_{L-1}}}{\tilde{w}_{g_1 \ldots g_{L-2}}} \right)^{-1} \cdot   \sum_{g_1} \ldots \sum_{g_{L-1} } \frac{\tilde{n}^0_{g_1 \ldots g_{L-1}}}{\tilde{w}_{g_1 \ldots g_{L-2}}}     \right] = N
\end{align}
Theorem \ref{theorem2} is hence proved from (\ref{e7.3})  and (\ref{e7.5}).
\subsection{Proof of Theorem 4}
In the simultaneous multi-way classification, $\hat{w}_{g_s} = \hat{w}_{g_{L=1}}$, the estimate of grouping effect in a hierarchical setup with one level of classification. Following arguments from the proof of theorem 2, we can show that $\hat{w}_{g_s}(\mathbf{P}) = \frac{n_{g_s} - R_{n_{g_s}}(\mathbf{P})+1}{N(1-\lambda)}$ is non-decreasing in each p-value. Hence the weights $\widehat{W}_i(\mathbf{P})$ as defined in equations (\ref{e3.11}) and (\ref{e3.12}) is non-decreasing in each p-value $P_i, i=1, \ldots, N$.

Note that 
\begin{align*}
E\left[\sum_{i \in I^0} \frac{1}{\widehat{W}_i(\mathbf{P}^{(-i)},0)}\right]  = E\left[\frac{1}{S}\sum_{s=1}^S \sum_{g_s = 1}^{M_s}\sum_{i \in I_{g_s}^0} \frac{1}{\hat{w}_{g_s}(\mathbf{P}^{(-i)},0)}\right] 
= \frac{1}{S}\sum_{s=1}^S E\left[\sum_{g_s = 1}^{M_s}\sum_{i \in I_{g_s}^0} \frac{1}{\hat{w}_{g_s}(\mathbf{P}^{(-i)},0)}\right]
\end{align*} where $\widehat{W}_i(\mathbf{P}^{(-i)},0)$ and $\hat{w}_{g_s}(\mathbf{P}^{(-i)},0)$ are respectively $\widehat{W}_i$ and $\hat{w}_{g_s}$ as functions of $\mathbf{P}^{(-i)} = \left\{P_1, \ldots, P_N\right\} \setminus \{P_i\}$ with $P_i=0$, $\forall i = 1, \ldots, N$. The last equality follows from the fact that the $S$ simulatenous classifications are assumed to be independent of each other. Following arguments from proof of theorem 2, it can be shown that
$$E\left[\sum_{g_s = 1}^{M_s}\sum_{i \in I_{g_s}^0} \frac{1}{\hat{w}_{g_s}(\mathbf{P}^{-i},0)}\right]\le N.$$ The proof of the theorem then follows from result \ref{result3}.
\bibliographystyle{Chicago}

\end{document}